\documentclass[twocolumn,apj]{emulateapj} 

\shorttitle{A geometrical height scale for sunspot penumbrae}
\shortauthors{Puschmann et al.}

\begin{document}


\title{A geometrical height scale for sunspot penumbrae}



\author{K. G. Puschmann\altaffilmark{1,2}, B. Ruiz Cobo\altaffilmark{1,2} and V. Mart\'\i nez Pillet\altaffilmark{1}}

\affil{(1) Instituto de Astrof\'isica de Canarias (IAC), E-38200 La Laguna, Tenerife, Spain\\
(2) Departamento de Astrof\'isica, Universidad de La Laguna (ULL), E-38205 La Laguna, Tenerife, Spain}
\email{kgp@iac.es, brc@iac.es, vmp@iac.es}






\begin{abstract}
Inversions of spectropolarimetric observations of penumbral filaments deliver
the stratification of different physical quantities in an optical depth scale.
 However, without establishing a geometrical height scale their
three-dimensional geometrical structure can not be derived. This is crucial in understanding 
the correct spatial variation of physical properties in the penumbral atmosphere and to provide insights into the mechanism capable of
explaining the observed penumbral brightness. The aim of this work is to determine
a global geometrical height scale in the penumbra by minimizing the divergence of
the magnetic field vector and the deviations from static equilibrium as imposed
by a force balance equation that includes pressure gradients, gravity and the
Lorentz force. Optical depth models are derived from the SIR
inversion of spectropolarimetric data of an active region observed with SOT 
on-board the Hinode satellite. We use a genetic algorithm to determine the
boundary condition for the inference of geometrical heights. The retrieved
geometrical height scale permits the evaluation of the Wilson depression at
each pixel and the correlation of physical quantities at each height. Our
results fit into the uncombed penumbral scenario, i.e., a penumbra
composed of flux tubes with channelled mass flow and with a weaker and more 
horizontal magnetic field as compared with the background field.
The ascending material is hotter and denser than their surroundings. 
 We do not find evidence of overturning convection or field free regions in the 
inner penumbral area analyzed. The penumbral brightness can be explained 
by the energy transfer of the ascending mass carried by the Evershed flow, 
if the physical quantities below $z$\,=\,-75\,km  are extrapolated from the 
results of the inversion.
\end{abstract}


\keywords{sun: sunspot, polarization, radiative transfer, methods: numerical}



\section{Introduction}

Our understanding of the structuring of the penumbral magnetic field has increased significantly over the last two decades. 
 Three seminal works were published back in 1993. Using filter magnetograms, \citet{titleetal93} found that the magnetic field 
inclination fluctuated by as much as $\pm$18 degrees (rms) on azimuthal paths around the spot center. \citet{litesetal93}, 
using spectropolarimetric data, identified the same fluctuations in the magnetic field, but this time in both inclination 
and field strength. The sense of the correlation found was that horizontal field lines tended to a have a weaker field strength. 
 In both works, the well known Evershed flow was observed to be concentrated in the horizontal component. Simultaneously,
 \citet{solankimontavon93} proposed a model, commonly referred to as the uncombed model, of vertically interlaced magnetic 
field components to explain the properties of the observed net circular polarization. 

The discovery of dark cored penumbral filaments by \citet{scharmeretal02}, once a resolution of $\sim$\,0\farcs1 was reached, led to the development of new models trying to explain the structure of penumbral filaments. Among them we can cite the flux pumping 
mechanism \citep{weissetal04, brummelletal08}, the MISMA model \citep{san05}, and the "gappy" 
model \citep{spruitscharmer06, scharmerspruit06}. For a detailed overview we refer to recent reviews of \citet{bellot09}, \citet{borrero09}, \citet{scharmer09}, \citet{sch09},  and \citet{tritschler09}. 

Many penumbral models \citep{sch98a,sch98b, marpill00, borreroetal05, borreroetal06, borrero07} share the geometry 
of the uncombed penumbral model: nearly horizontal magnetic flux tubes embedded in a stronger and more vertical 
background field. Throughout the article we will refer them as the uncombed scenario. In the frame of this 
scenario, several observational properties of sunspot penumbrae, like the filamentary structure, the strong Evershed outflow, 
and the uncombed geometry can be easily explained. Besides, it explains the properties of the observed net circular 
polarization \citep{sch02, schetal02, muelleretal02, muelleretal06}. Recently the uncombed scenario has been modified, 
see \citet{borrerosolanki10}, to make it compatible with a possible existence of convective motions inside penumbral 
filaments as pointed out by recent observations \citep{bellogonzalezetal05, ichimotoetal07b, zakharovetal08}. One open 
question in the uncombed scenario is how the penumbral continuum intensity can be as large as 75\% of the normal 
quiet Sun brightness \citep[see, e.g.,][]{solankirueedi03, sch03, spruitscharmer06}. 
 However, it is interesting to note that the simulations by \citet{Ruizcobobellotrubio08} can explain up 
to 50 \% of the quiet Sun intensity if the Evershed flow consists of horizontal flow channels with a length of 
several Mm that bring hot material from deeper layers. Indeed, the problem of the penumbral heating  was a strong 
motivation for \cite{scharmerspruit06} and \cite{spruitscharmer06} to propose the gappy penumbral model, in 
which field-free overturning convection occurs inside the penumbra and efficiently heats its atmosphere. However, 
this model has not yet been confronted to spectropolarimetric observations. In particular, it is unclear if this 
model is able to reproduce the observed net circular polarization. However, for 
that to happen, the magnetic field inside the filaments should be around 1000 G, which seems to be incompatible 
with the concept of a field-free gap \citep{borrerosolanki10}. Recently, \cite{scharmer08} proposed the possibility of strongly reduced (although non-zero) magnetic field inside the field-free gap. These modifications of both 
scenarios, uncombed (including convection inside flux tubes) and gappy (including significant magnetic 
field inside the field-free gaps), could in the end produce similar observational effects, although they are based on 
different physical mechanisms.

Simulations based on radiative magneto-convection in inclined magnetic fields can help to shed light on open questions. 
 First attempts to simulate penumbral structure in small slab-like sections of sunspots \citep{heinemann07, rempeletal09a} resulted 
in rather narrow penumbral regions. Recent numerical simulations by \cite{rempeletal09b} seem to reproduce many observational 
properties of penumbral filaments: e.g. the filamentary structure of the penumbra, the Evershed flow, some supersonic 
velocities \citep[although the observed ones are certainly more complex, see e.g.][]{bellotrubioetal04, ichimotoetal07a, marpill09}, 
the correlation between inclination and strength of the magnetic field, and the penumbral brightness. Nevertheless, the resulting 
penumbra is still somewhat immature, revealing too short and too fragmented filaments. Dark lanes along bright filaments observationally found by \citet{scharmeretal02} form only occasionally. At present, these simulations do not allow a direct comparison with 
spectropolarimetric observations, since up to our knowledge no synthesized spectra are available. Surprisingly, 
the simulated penumbrae share some properties with both the uncombed and the gappy scenarios: i.e. convective motions 
inside zones harboring nearly horizontal and relatively strong (around 1kG) magnetic fields. 
 The vertical slices across a penumbral filament taken from a snapshot of Rempel et al's simulations and presented in Figure 1 
of \cite{borrero09} point to the co-existence of significant magnetic fields {\em and} overturning velocities.

Observationally, 3D visualizations of penumbral filaments can only be obtained from inversions of 
spectropolarimetric data \citep[in the absence of robust results from local helioseismology,][]{gizon09}. 
 While Milne-Eddington inversions only yield the physical parameters at an average optical depth  \citep{sanetal96}, 
more sophisticated inversion codes like SIR \citep{ruizcobodeltoro92} or SPINOR \citep{frutigeretal00} provide 
the stratification of physical parameters in an optical depth scale. Several 3D optical depth models have been 
published in the recent past \citep[see, e.g.,][]{westend01a, westend01b, mathewetal03, monica05, beck06, beck08}. 

The highest spatial resolution input data for this endeavor are currently provided by the CRisp 
Imaging SpectroPolarimeter \citep[CRISP,][]{scharmer06} and by the Hinode/SP spectropolarimeter 
(Lites et al. 2001, see also Tsuneta et al. 2008). In the near future the Gregor Fabry Perot 
interferometer \citep[GFPI, see e.g.][]{pus06,pus07} and its full Stokes polarimeter 
\citep{bellokneer08,balthasaretal09} $\bf may$ reveal spectropolarimetric data at even higher spatial resolution 
once Gregor \citep{volkmeretal07} is in operation. The ground based 2D observations take advantage of 
image reconstruction techniques 
\citep[see e.g.][]{loefdahl02, vannoortetal05, puschmannsailer06, denkeretal07, bellokneer08, woegervdl08} to improve 
spatial resolution and to reduce spatial crosstalk from variable seeing.

\citet{jurcak07} and \citet{jrcakbellotrubio08} have carried out 
inversions of Hinode data and found an atmosphere that is basically consistent with the uncombed scenario: 
horizontal channels carrying the Evershed flow are substantially magnetized. These channels appear at large 
optical depths, but no information about geometrical heights is obtained. 

Given the highly fluctuating densities 
that are expected in the penumbra, the interpretation of an optical
 depth scale as a good approach for geometrical heights can be misleading and a reliable transformation
 to geometrical heights is mandatory. Besides, to establish a geometric height scale is important for the determination of 
the electrical current vector $\vec{J}$ that is crucial for the determination of ohmic energy dissipation, and 
for a test of the reliability of {\it ab initio} MHD simulations. While traditionally in quiet Sun the transformation from an 
optical depth scale to a geometric one has been done by assuming hydrostatic equilibrium \citep[see, for 
example,][]{pus05}, this is not justified in the magnetized penumbra. The force balance in this case must 
include magnetic forces which require the calculation of horizontal {\em and} vertical spatial derivatives of the 
magnetic field. 
  
An absolute geometrical height scale for penumbrae has been derived by \cite{san05} from a 
spectropolarimetric inversion under the MISMA hypothesis by imposing equal total pressure 
between adjacent pixels, although neglecting the magnetic tension in the Lorentz force. \cite{carrolkopf08} 
obtained the stratification of physical quantities in geometrical height applying a neural network 
inversion technique based on MHD simulations of quiet Sun. The application of this technique on sunspot penumbrae might be straight forward, 
using e.g. the simulations of \cite{rempeletal09b} as input.


The main objective of the work presented here is to infer a 3D 
geometrical model of the penumbra by imposing the solenoidality of the magnetic field and the 
dynamic equilibrium including the Lorentz force. This model allows the determination of the Wilson depression at
each pixel, the correlation of physical quantities at each height, and the evaluation of the energy budget transferred by mass motions.

\section{Observations}
The active region AR 10953 near solar disk center was observed using the spectropolarimeter (SP) of the Solar Optical
Telescope (SOT) on-board the Hinode spacecraft on the 1$^{\rm st}$ of May 2007, between 10:46 and
12:25 UT. The region was scanned with a step size of 0\farcs148 and a slit width corresponding to 0\farcs158, recording the full Stokes
vector of the pair of the neutral iron lines at 630\,nm with a spectral
sampling of 21.53 m\AA. The spatial resolution was about 0\farcs32.

   \begin{figure}
   \centering
   \includegraphics[width=10cm]{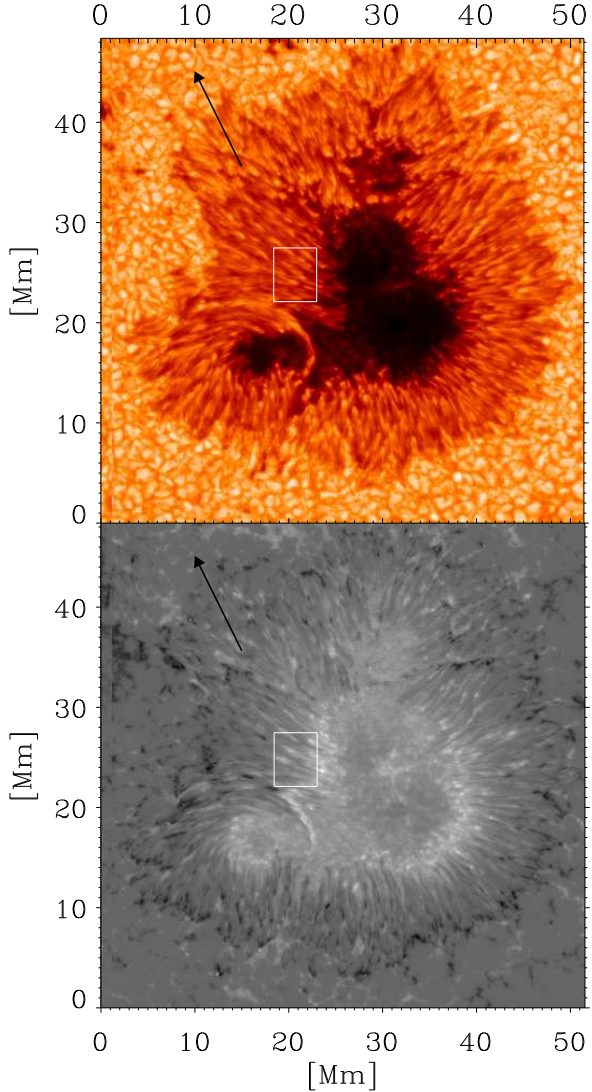}
   \caption{Upper panel: Continuum intensity at 630.2\,nm of a fragment of the active region AR 10953 observed 
by SOT on-board of the Hinode spacecraft. The white rectangle marks the area under study in the 
present paper. 
The arrow indicates the direction to disk center. The slit is vertically oriented.
Bottom panel: Stokes V at +286 m\AA~ from the line center.
}
   \label{Fig1}
   \end{figure}

The integration time was 4.8\,s, resulting in a noise level of approximately 
1.2\,$\times$\,10$^{-3}$. The wavelength calibration was done assuming
that the average umbral profile does not exhibit velocities. 

For reasons of simplicity this article centers on a section of the inner penumbra 
of the large sunspot presented in Figure~\ref{Fig1}. The analyzed area is marked by a white rectangle 
centered at an heliocentric angle $\theta$\,=\,4.63$^{\circ}$. The arrow indicates the direction to 
disk center. In figure coordinates, the disk center would be at $X$\,=\,-5.5\,Mm and $Y$\,=\,74\,Mm.
 This area has been chosen because it consists of homogeneous and radial aligned filaments.
In the bottom panel of Figure~\ref{Fig1} we show the map of Stokes V at +286 m\AA~ from the line center. 
In this map, the black points correspond to pixels in which the magnetic field shows a reverse polarity 
and harbors strong downflows \citep{ichimotoetal07a}. The area analyzed in this paper does not include
such points. Nevertheless, provided that these strong downflows are mainly located in 
the mid and outer penumbra, the results of this paper can be considered representative for the inner penumbra.
 In a posterior paper we will apply the method described in this article to the entire sunspot.

\section{Inversion procedure}
To derive the physical parameters of the solar atmosphere as a function of continuum optical depth, i.e. temperature $T(\tau$), magnetic field strength $B(\tau)$, field inclination $\gamma(\tau)$, field azimuth $\phi(\tau)$, and line of sight velocity $v_{\rm los}(\tau)$, the SIR 
inversion code \citep{ruizcobodeltoro92} was applied on the spectropolarimetric data set. In the region of the inner center-side penumbra under study, all Stokes profiles are slightly asymmetric, are not multi-lobed and do not show strong displacements. Consequently, there is not sufficient information to obtain a two-component atmosphere unambiguously. Therefore, we use an one-component inversion to obtain the stratification of plasma parameters in each pixel. The same procedure has already been applied on sunspot penumbral data observed with the Hinode/SP by e.g. \citet{jrcakbellotrubio08}. The spectropolarimeter on-board Hinode allows us to distinguish the fine structure of the penumbra with a resolution of 0\farcs3. The dark cored penumbral filaments discovered by \citet{scharmeretal02}, with a typical size less than 0\farcs1, could indicate the existence of smaller physical structures, but could also be generated by temperature differences present in iso-tau layers above large structures like flux tubes \citep{Ruizcobobellotrubio08} or field-free gaps \citep{spruitscharmer06}. In any case, our results have to be interpreted as the averaged properties of structures on a scale of 0\farcs3.

The free physical parameters are calculated by SIR at different optical depth points (nodes).  We have chosen 5 nodes in $T(\tau)$, 3 nodes in $B(\tau)$ and $v_{\rm los}(\tau)$, and 2 nodes in $\gamma(\tau)$ and $\phi(\tau)$. The final temperature stratification is
obtained by spline interpolation. We have used a parabolic interpolation for
$B(\tau)$ and $v_{\rm los}(\tau)$ and linear interpolation in case of
$\gamma(\tau)$ and $\phi(\tau)$.  We have neither considered microturbulent
velocities nor stray light contamination.
 The final synthesized profiles have
been convolved with a macroturbulent velocity $v_{\rm mac}$ as an additional
free parameter of the inversion.
   \begin{figure}
   \centering
   \includegraphics[width=9cm]{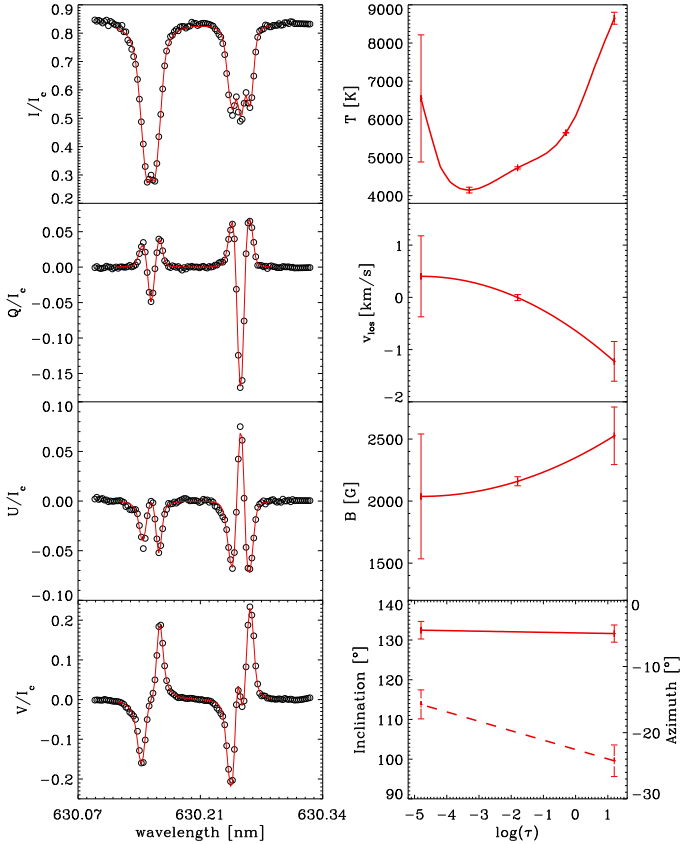}
   \caption{Left column: Normalized observed Stokes $I$, $Q$, $U$ and
   $V$ profiles of one pixel (circles); The red solid lines correspond to the best fits of synthesized profiles. 
    Right column: Corresponding stratification of
   temperature $T$, line-of-sight velocity $v_{\rm los}$, magnetic field
   strength $B$, field inclination $\gamma$ and field azimuth $\phi$ (dashed
   line) {\it vs.} logarithm of continuum optical depth at 500 nm $\log(\tau)$;
   over-plotted are the error bars calculated at the different nodes.}
   \label{Fig2}
   \end{figure}

In Figure~\ref{Fig2}, we present the results of the inversion at an arbitrarily
chosen pixel. Note that the synthesized spectral lines even reproduce the
asymmetries present in the observational Stokes profiles (left four panels in Figure~\ref{Fig2}).
 In the right four panels of this figure, we present the stratification of $T$,
$v_{\rm los}$, $B$, $\gamma$, and $\phi$ versus the continuum optical depth.
 The error bars calculated at the different nodes are also shown. The sunspot
has negative polarity ($\gamma>90^{\circ}$). We do not perform any correction
of the azimuth to solve the 180$^{\circ}$ ambiguity, since all resulting
azimuth values are in the range between \mbox{-90}$^{\circ}$ and 90$^{\circ}$ showing
a smooth spatial variation across the field of view (FOV).

\section{Determination of a geometrical height scale}

\begin{figure}
   \centering
   \includegraphics[width=9cm]{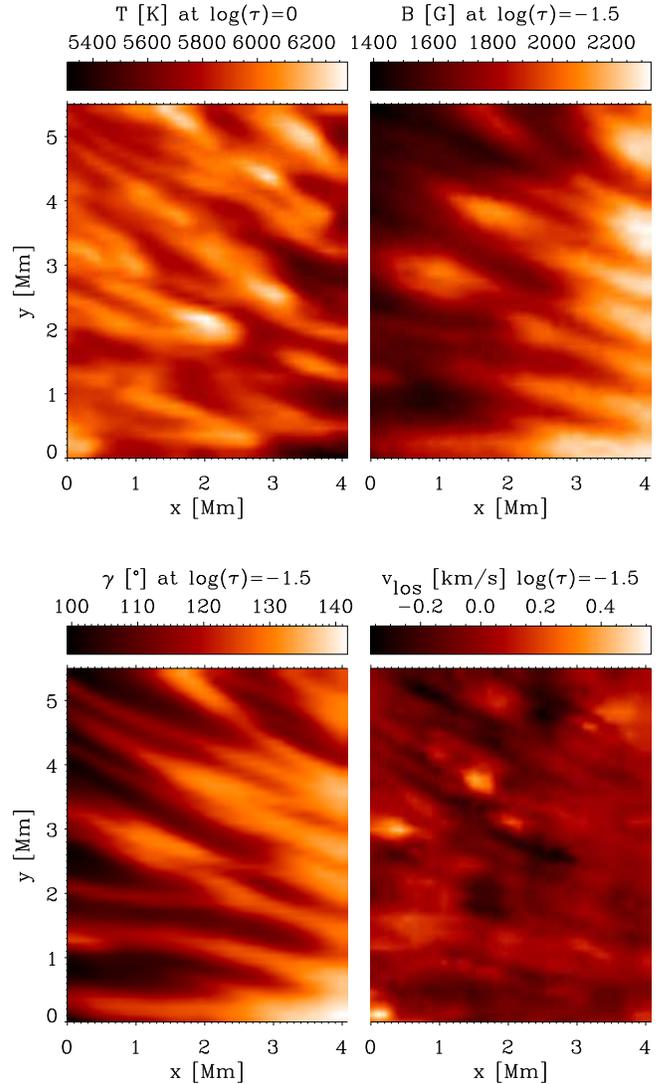}
      \caption{Maps of temperature $T$ at $\log(\tau)$\,=\,0, and maps of
      magnetic field strength $B$, field inclination $\gamma$, and line of
      sight velocity $v_{\rm los}$ at $\log(\tau)$\,=\,-1.5. Negative and positive velocities
correspond to upflows and downflows, respectively.}
   \label{Fig2bis}
   \end{figure}

The inversion method delivers for each pixel the stratification of an
atmospheric model {\it vs.} continuum optical depth, i.e. we obtain
$\vec{B}(x,y,\tau)$, $T(x,y,\tau)$, etc\dots (see Figure~\ref{Fig2bis}). Taking into account 
that the region under investigation is not exactly placed at disk center ($\mu$\,=\,0.997), 
our $XY$-plane is not parallel to the local solar surface. A transformation of the reference system can only be 
applied {\it a posteriori}, once the $z$-scale has been obtained. Besides, it is impossible to perform this 
transformation without the knowledge of horizontal velocities. Consequently, we consider the $z$-axis 
to be parallel to the line-of-sight direction throughout this article. We will term the $z$-axis as the 
"vertical" and accordingly the perpendicular plane as the "horizontal" plane.

A geometrical height scale $z(x,y,\tau)$ can be obtained by integrating

\begin{equation}
dz=-\frac{1}{\kappa \rho}\, d\tau \,\,.
\label{optical depth}
\end{equation}

For the integration of equation~(\ref{optical depth}), three ingredients are needed: $\kappa$ (continuum
absorption coefficient per gram), $\rho$ (mass density) and the boundary
condition $Z_{\rm W}=-z(\tau=1)$ (Wilson depression)\begin{footnote}
{We introduce a change of sign to be consistent with the usual definition of the Wilson depression, 
i.e. positive values for depressed $\tau$\,=\,1 layers.}\end{footnote}. The first ingredient, 
$\kappa(x,y,\tau)$, is evaluated by SIR from
$T(x,y,\tau)$, the gas pressure ($P_{\rm g}$) and the abundances.
 However, as we will see below, $\kappa$ is nearly independent of $P_{\rm g}$.
 The stratification of $P_{\rm g}$ and $\kappa$ are obtained under the assumption of hydrostatic equilibrium (HE),
i.e., by the integration of
\begin{equation}
dP_{\rm g}/d\tau= g/\kappa\,\,,
\label{pgtau}
\end{equation}
where $g$ represents gravity. In each step of the integration (i.e.~at each $\tau$) $\kappa$ and $P_{\rm g}$ are found iteratively such that both variables are compatible with the ionization equilibrium in LTE (Saha equation). For the gas
pressure boundary condition at the top layer ($\log\tau$\,=\,-4.8) we used 655 dyn/cm$^2$ \citep[taken from][]{vernazzaetal81}.
 Solving the HE (equation~(\ref{pgtau})) with boundary conditions differing by two orders of magnitude ($P_{\rm g}$\,$(\log\tau$\,=\,-4.8$)$\,=\,10\,dyn/cm$^{2}$ and $P_{\rm g}$\,$(\log\tau$\,=\,-4.8$)$\,=\,1000\,dyn/cm$^{2}$) the resulting stratification of the absorption coefficients ($\kappa_{10}(\tau)$ and $\kappa_{1000}(\tau)$) is practically invariant. The ratio between $\kappa_{1000}(\tau)$ and $\kappa_{10}(\tau)$ approximately follows the empirical law: 
\begin{equation}
\frac{\kappa_{1000}(\tau)}{\kappa_{10}(\tau)}\sim1-\frac{5\cdot10^{-5}}{\tau}\,\, ,
\label{kappa}
\end{equation}
i.e. at $\log\tau$\,$>$\,-2 the difference is smaller than 0.5\,\%.

The pressure stratification is used for the calculation of the second
ingredient, the density, using the equation of state for an ideal gas considering
partial ionization.

By choosing an arbitrary Wilson depression $Z_{\rm W}$ at all pixels, a geometric height scale
$z(x,y,\tau)$ can be constructed after integration of equation~(\ref{optical
depth}) (which would be a HE height scale).  
 $\vec{B}(x,y,z)$ maps are then obtained by the interpolation of the
$\vec{B}(x,y,\tau)$ maps resulting from the SIR inversion. Such
$\vec{B}(x,y,z)$ maps have a divergence different from zero. Furthermore,
this model would not be in mechanical equilibrium. An optimum choice of $Z_{\rm
W}(x,y)$ should minimize at a given height both the divergence of the magnetic
field $\nabla\cdot\vec{B}$ and the error in the equation of motion.

Neglecting viscosity, the equation of motion can be written as
\begin{equation}
\rho \frac{d\vec{v}}{dt}=\vec{J}\times\vec{B}+\rho\,\vec{g}-\nabla P_{g}\,\, ,
\label{motioneq}
\end{equation}
where $\vec{v}$ stands for velocity and $\vec{J}$ for the current density.  The
acceleration on the left hand side of equation~(\ref{motioneq}) can be decomposed in a
temporal derivative and an advective term
\begin{equation}
\frac{d\vec{v}}{dt}=\frac{\partial\vec{v}}{\partial t}+(\vec{v}\cdot\nabla)\vec{v}\,\,.
\label{derivada}
\end{equation}
Due to the extended lifetime of penumbral filaments of $\sim$\,1 hour \citep[see e.g.][]{suetterlinetal04, langhansetal07}
compared to the sound travel time across a filament, their velocity can be
considered stationary and we thus neglect the partial temporal derivative $\frac{\partial\vec{v}}{\partial t}$
in equation~(\ref{derivada}). Since the photospheric magnetic field is nearly frozen into the plasma, we can assume, as an order of magnitude estimation, that the material moves along the field lines and derive $\vec{v}$ from
$v_{\rm z}$\,=\,$-v_{\rm los}$, $\gamma$, and $\phi$. Note that positive $v_{\rm los}$ corresponds to a red-shift while positive $v_{\rm z}$ are upflows. 

\begin{figure}
   \centering
   \includegraphics[angle=0,width=7cm]{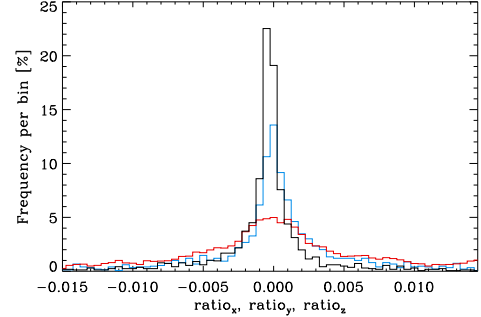}
      \caption{Histograms of the ratio between the advective term and the dynamic term without considering the Lorentz force for
      each of the three components evaluated at a height of 200\,km in the
      penumbral region presented in Figure~\ref{Fig1}: ratio$_{\rm x}$ (black),
      ratio$_{\rm y}$ (red), ratio$_{\rm z}$ (blue). Binsize of 5\,$\times$\,10$^{-4}$.}
         \label{histdynamico}
   \end{figure}

We also negelect the dynamic term $(\vec{v}\cdot\nabla)\vec{v}$, since it can be shown to be much smaller than the advective term.
 In Figure~\ref{histdynamico} we plot the histogram of the errors introduced by neglecting the dynamic term: ${\rm ratio}_{\rm x}$, ${\rm ratio}_{\rm y}$, ${\rm ratio}_{\rm z}$, which represent the ratio between the advective term (the second term on the right hand side of equation~(\ref{derivada})) and the dynamic term (second and third terms on the right hand side of equation~(\ref{motioneq})) without considering the Lorentz force for each of the three components:
\begin{eqnarray}
\label{xyz}
{\rm ratio}_{\rm x} & = & -\rho\frac{(\vec{v}\cdot\nabla)v_{\rm x}}{\partial{P_{g}}/\partial{x}} \nonumber \\
{\rm ratio}_{\rm y} & = & -\rho\frac{(\vec{v}\cdot\nabla)v_{\rm y}}{\partial{P_{g}}/\partial{y}}  \\
{\rm ratio}_{\rm z} & = & -\rho\frac{(\vec{v}\cdot\nabla)v_{\rm z}}{\partial{P_{g}}/\partial{z} + \rho g}\,\,. \nonumber
\end{eqnarray}
The magnitudes of equation~(\ref{xyz}) have been evaluated at the reference height of 200\,km where the minimization procedure will be applied. The neglection of the advective term is justified since most of the values are below 1\%. 

Let us define then the residual force $\vec{F}$ to be
\begin{equation}
\vec{F}=\vec{J}\times\vec{B}+\rho\,\vec{g}-\nabla P_{g}\,\, ,
\label{appmotioneq}
\end{equation}
with $P_{g}$ inferred from equation~(\ref{pgtau}).
In our approximation $\vec{F}$ is zero if the equation of motion is fulfilled. To ensure the physical meaningfulness of the solution also
$\nabla\cdot\vec{B}$ must be zero.

We define a merit function as
\begin{equation}
\chi^{2}=\sum_{pixels}{w_{1}(F_{\rm x}^{2}+F_{\rm y}^{2}+F_{\rm z}^{2})+w_{2}(\nabla\cdot\vec{B})^{2}}\,\,.
\label{meritfunction}
\end{equation}
where $w_{\rm 1}$, $w_{\rm 2}$ are coefficients introduced to adequately weight
the contribution of both $\nabla\cdot\vec{B}$ and $\vec{F}$. By introducing
vertical displacements of the atmospheric models at each pixel, we minimize 
the merit function of equation~(\ref{meritfunction}). Therefore, we used a 
genetic algorithm that iteratively changes 
an array $D_{\rm z}(x,y)$ containing the displacements of 
the atmospheric models at each pixel. The merit function is evaluated at the 200\,km height level. At this layer the response function of Stokes V of the visible lines to the magnetic field strength reaches its maximum and consequently the uncertainties of the inverted magnetic field are minimal. Besides, at this layer the error introduced by neglecting the acceleration term in equation~(\ref{derivada}) is expected to be 
smaller than at deeper photospheric layers since, in sunspot penumbrae, the velocity field is confined to lower layers. 

\begin{figure}
   \centering
   \includegraphics[angle=0,width=9cm]{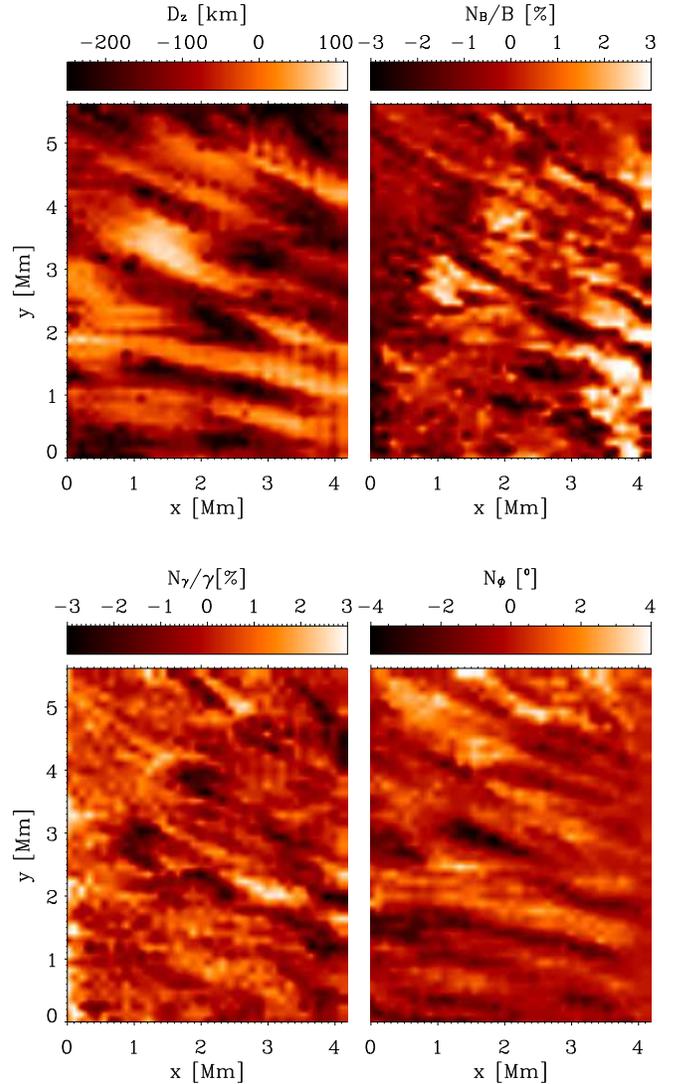}
   \caption{Solutions of the genetic algorithm: Displacement of the
   $z$\,=\,200\,km layer $D_{\rm z}(x,y)$ and noise functions $N_{B}$,
   $N_{\gamma}$, and $N_{\phi}$. See text for details. We do not plot relative
   $N_{\phi}$ values because ${\phi}$ reaches null values at some pixels.}
  \label{desplazamientos}
\end{figure}

  Owing the uncertainties of the inversion method, the obtained stratification of the magnetic field is neither solenoidal nor does it satisfy the equation of motion. Thus, for the evaluation of $\chi^{2}$ we use slightly modified values for the magnetic field
vector inside its uncertainties, i.e.~ we define
$B'(x,y,z)$\,=\,$B(x,y,z)$\,+\,$N_{B}(x,y)$,
$\gamma'(x,y,z)$\,=\,$\gamma(x,y,z)$\,+\,$N_{\gamma}(x,y)$, and
$\phi'(x,y,z)$\,=\ \,$\phi(x,y,z)$\,+\,$N_{\phi}(x,y)$, with absolute values
of noise functions $N_{B}(x,y)$, $ N_{\gamma}(x,y)$, $ N_{\phi}(x,y)$ smaller
than the error uncertainties of the respective parameters. The resulting
synthesized Stokes profiles, considering $B'$, $\gamma'$ and $\phi'$, are still
compatible with the observed Stokes profiles. In summary, the solution found by
the genetic algorithm is constituted by the optimum values of $D_{\rm z}(x,y)$,
$N_{B}(x,y)$, $N_{\gamma}(x,y)$, and $N_{\phi}(x,y)$.  We have performed
several realizations of the genetic algorithm to determine the optimum values
of the weights $w_{\rm 1}$, $w_{\rm 2}$. If we set $w_{\rm 1}$ to zero (i.e.
just minimizing $\nabla\cdot\vec{B}$) many different solutions are reached; with
$w_{\rm 2}$ equal to zero (i.e., minimizing the components of $\vec{F}$)
the solution reduces in particular the $\nabla\cdot\vec{B}$ values,
however not satisfactorily.  The best solution is reached setting $w_{\rm
1}$ and $w_{\rm 2}$ such that both addends in equation~(\ref{meritfunction}) contribute
in an equal manner ($w_{\rm 2}$\,=\,3.3$\cdot$10$^{5}\cdot w_{\rm 1}$, using
cgs units). In each realization, the code produces slightly different results
with an approximately Gaussian distribution around a mean value at each pixel.
 Therefore we adopted the average of 20 individual realizations as the final solution.
 In the upper left panel of Figure~\ref{desplazamientos} we present the resulting
displacement $D_{\rm z}(x,y)$. The displacements span a range between -250\,km and
100\,km and show a spatial distribution well correlated with the $x$\,-\,$y$
maps of several physical quantities, e.g.~ $T$, $B$, or $\gamma$ (see Figure~\ref{Fig2bis}). 
 Negative numbers in this scale correspond to deeper layers. The zero height of
the scale has been arbitrarily set at one of the pixels analyzed; its comparison
with an absolute height scale in the quiet Sun photosphere is discussed in Sect. 5.
 The relative noise $N_{B}(x,y)$, $N_{\gamma}(x,y)$, and
$N_{\phi}(x,y)$ is presented in the upper right and lower panels of Figure~\ref{desplazamientos}. Note that the introduced noise functions account for errors below 3\% in $B$ and $\gamma$, and below 4$^{\circ}$ in case of $\phi$.

\begin{figure}
   \centering
   \includegraphics[angle=0,width=9cm]{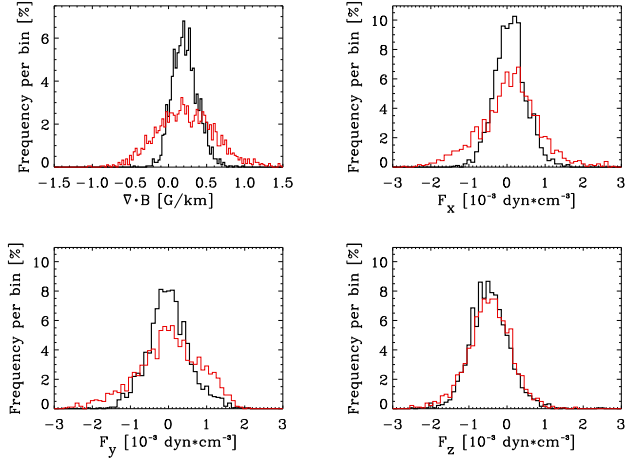}
   \caption{Histograms of $\nabla\cdot\vec{B}$ and the three components of the residual force $\vec F$ at $z$\,=\,200\,km before (red lines) and after (black lines) the application of the genetic algorithm. Binsize of 0.025
   ($\nabla\cdot\vec{B}$) and 10$^{-4}$ (force components), respectively.}
   \label{div_F}
\end{figure}
\begin{figure}
   \centering
   \includegraphics[angle=0,width=9cm]{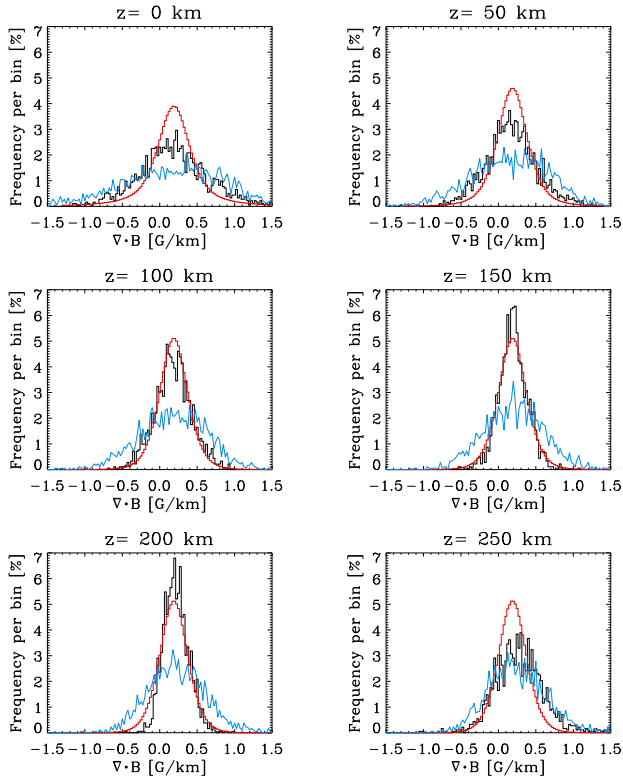}
   \caption{Black: Histograms of the divergence of the magnetic field strength at six
   different layers after the integration of the vertical component of the
   equation of motion. Red: Mean histograms of the divergence obtained from a 
   random Gaussian field distribution. The error histograms have been shifted by
   0.2\,G/km. Blue: Histograms obtained before the application of the genetic algorithm. 
    Binsize of 0.025}
   \label{divcapas}
\end{figure}

After introducing $D_{\rm z}(x,y)$, $N_{B}(x,y)$, $N_{\gamma}(x,y)$ and
$N_{\phi}(x,y)$ and interpolating to a common $z$\,-\,scale we can assume that
the layer at 200\,km approximately satisfies both, $\nabla\cdot\vec{B}$ and the
equation of motion, although the pressure stratification for each pixel continues
being the HE one. We can obtain now a more accurate $P_{\rm
g}$\,-\,stratification by the integration of the $z$\,-\,component of  equation~(\ref
{appmotioneq}). However, when the $P_{\rm g}$\,-\,stratification changes, the
$z$\,-\,scale is also modified. Provided that $T(\tau)$ and $B(\tau)$ are
invariant, it is easier to integrate this equation in terms of optical depth.
 We can rewrite the vertical component of  equation~(\ref{appmotioneq}) in terms of $\tau$, setting $\vec{F}$\,=\,0:
\begin{equation}
\frac{dP_{g}}{d\tau}=\frac{g}{\kappa}-\frac{R T}{\kappa\mu P_{g}}{(\vec{J}\times\vec{B})}_{\rm z}\,\, .
\label{motioneqtau}
\end{equation}
After the integration of this equation, e.g. by Runge-Kutta, we obtain $P_{\rm g}$. 
 From the equation of state we evaluate $\rho$ and subsequently the new $z$\,-\,scale 
from equation~(\ref{optical depth}). The newly obtained $P_{g}$ values produce 
slightly modified stratification and the whole procedure starting with 
equation~(\ref{appmotioneq}) is again repeated. The modifications are minimal and 
convergence is achieved after two iterations. Figure~\ref{div_F} shows the histograms 
of the divergence of the magnetic field and of the three components of the residual force 
before and after the minimization. The process clearly improves $\nabla\cdot\vec{B}$ 
(changing the standard deviation of its histogram, $\sigma$, from 0.38 to 0.16 G/km), 
$F_{\rm x}$ ($\sigma$ from 8$\cdot$10$^{-4}$ to 4$\cdot$10$^{-4} \rm{dyn/cm}^3$), and $F_{\rm y}$ ($\sigma$ from 8$\cdot$10$^{-4}$ 
to 6$\cdot$10$^{-4} \rm{dyn/cm}^3$) and slightly diminishes $F_{\rm z}$ ($\sigma$ from 6$\cdot$10$^{-4}$ 
to 5$\cdot$10$^{-4} \rm{dyn/cm}^3$).

The models at all pixels were then interpolated to a common global $z$\,-\,scale. In
Figure~\ref{divcapas} we present the resulting histograms of $\nabla\cdot\vec{B}$ at
six different layers. To check the significance of the results, we evaluated
histograms of the simulated magnetic field distributions which deviate from
solenoidal fields only by Gaussian noise, with a sigma equal to the  estimated
errors of the field components at each layer. We plot the mean
histograms of 1000 realizations shifted by 0.2\,G/km as red lines.  Owing to the resulting
coincidence between the histograms of both $\nabla\cdot\vec{B}$ and the errors, we
conclude that in the range 50\,km to 200\,km, the residual value of $\nabla\cdot\vec{B}$ can be
entirely attributed to noise except for a systematic offset of 0.2\,G/km. At lower and higher layers, not all $\nabla\cdot\vec{B}$ values can be
regarded as noise. The blue lines represent $\nabla\cdot\vec{B}$ calculated with a
$z$\,-\,scale obtained in hydrostatic equilibrium and considering as boundary
condition $Z_{\rm W}$\,=\,0 and
$P_{g}(\log(\tau)$\,=\,-4.8$)$\,=\,655\,dyn\,cm$^{-2}$. Note that at layers
below 250\,km the divergence of the magnetic field vector improves considerably
after applying the genetic algorithm.

\begin{figure}
   \centering
   \includegraphics[angle=0,width=9cm]{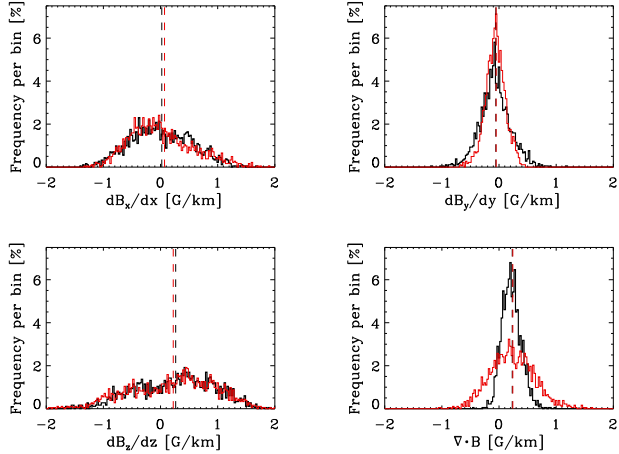}
   \caption{Horizontal (top) and vertical (bottom left)
   derivatives both before (red) and after (black) the minimization by the genetic algorithm. Vertical dashed lines
   correspond to the mean values of the distributions. The bottom right panel shows the sum of the three derivatives and is the same histogram as shown in Figure~\ref{div_F} at other scale.
   }
   \label{compo}
\end{figure}

What remains elusive is the origin of the systematic 0.2\,G/km offset. To
understand how this value is distributed among the three components of 
$\nabla\cdot\vec{B}$,  the horizontal derivatives ${\partial
B_{\rm x}\over{\partial x}}$, ${\partial B_{\rm y}\over{\partial y}}$ and the vertical
derivative ${\partial B_{\rm z}\over{\partial z}}$, in Figure~\ref{compo} we show these components separately (both before and after the minimization of the merit function by the genetic algorithm). 
 It turns out to be somewhat surprising that the individual histograms of the
derivatives in the three Cartesian directions do not change much with the application of the
minimization algorithm. It is only their addition in the form of the divergence
that is effectively minimized by the algorithm.  But in this figure, we
readily see that the offset of 0.2\,G/km originates entirely from the vertical
derivative.  This should be no surprise. Already \citet{westend01a}, who
performed a similar analysis to the one done in this work but without computing a
geometrical height scale, found an excess in the vertical derivative after
inverting data obtained with the Advanced Stokes Polarimeter \citep[see][]{litesetal93}.
 These data had a spatial resolution at least three times worse than that used in
our work. \citet{westend01a} attributed the origin of this excess to the unresolved nature
of the structuring of the penumbral magnetic field. Interestingly, they reported on a mean
excess 5-10 times larger than the 0.2\,G/km found here. 
 The question of how the vertical derivative can be made compatible
with the null divergence condition was debated by several authors. 
 \citet{san98} suggested that a microstructuring of the penumbral magnetic field in ranges around 
1-10\,km is able to solve this problem. Following \citet{marpill00}, the azimuthal components generated 
by the background field wrapping around penumbral flux tubes cancel out in the resolution 
element leading to the small horizontal gradients observed \citep[see also][]{borrero08}.
 In Figure 1 of \citet{san98} one finds the magnitude of the
horizontal derivatives that enter the calculation of the divergence 
for observations with resolution worse than 0\farcs2. The values cited
fluctuate with typical magnitudes of 0.15\,G/km, clearly smaller than the vertical gradients estimated
with 1 arcsec resolution spectropolarimetry. The situation in our present
analysis has improved considerably as now the horizontal derivatives fluctuate around values that are of the same order of magnitude
than the vertical one (see Figure~\ref{compo}). However, the excess of 0.2\,G/km still shows the presence of
unresolved components that could be present either in the vertical or in the horizontal
directions.

\begin{figure}
   \centering
   \includegraphics[angle=0,width=9cm]{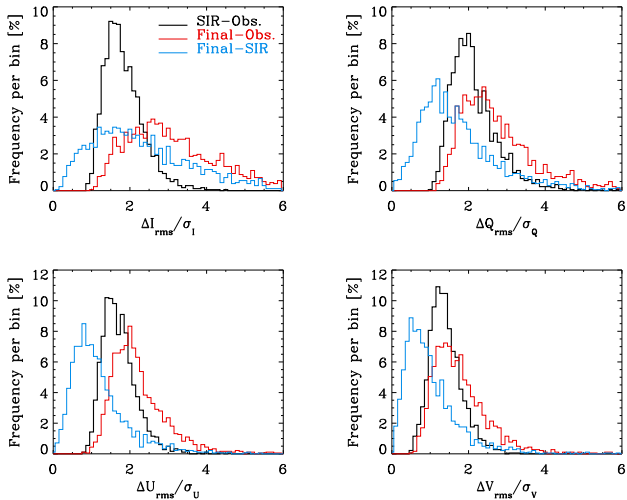}
   \caption{Histograms of the rms differences between the inverted (SIR) and
   the observed (Obs.) Stokes profiles in units of the standard deviation of
   the noise for the respective Stokes parameter (black). Histograms of the rms differences between the Stokes profiles
   synthesized from the final models (Final) and the observed profiles (red) and between the final and the SIR profiles (blue).}
   \label{fig_Rolf}
\vspace{0.5truecm}
\end{figure}

In order to estimate the compatibility of the resulting models with the
observed Stokes profiles
$(\mbox{I}^{obs},\mbox{Q}^{obs},\mbox{U}^{obs},\mbox{V}^{obs})$, we have
synthesized the Stokes profiles
$(\mbox{I}^{final},\mbox{Q}^{final},\mbox{U}^{final},\mbox{V}^{final})$ from
the final model atmospheres. In the upper left panel of Figure~\ref{fig_Rolf}, we
present the histogram of the rms difference
$\Delta$I$_{\mbox{rms}}=\sqrt{\Sigma({\mbox{I}}^{final}-{\mbox{I}}^{obs})^2}$
between the final and the observed profiles (red line). The histogram of rms
differences between the best fit found by SIR (labelled in the figure as SIR)
and the observed profiles are plotted in black. This rms is precisely the
quantity minimized by the inversion method. The histogram of rms differences
between the final and the SIR profiles (i.e. the error introduced by the
modification of the model through the application of the present technique) is
presented in blue. The corresponding histograms for the other Stokes
parameters are shown in the remaining panels of this figure. Note that the
rms differences between the SIR and the observed profiles peak around
1.5\,$\sigma$ -- 2\,$\sigma$ (where $\sigma$ is the standard deviation of the
noise in the respective observed Stokes parameter), while the additional noise
introduced by the present technique has a distribution that peaks (in case of
Q, U, and V) at around 1$\sigma$. This is compatible with the fact that the noise
functions, $N_{B}(x,y)$, $ N_{\gamma}(x,y)$, $ N_{\phi}(x,y)$ introduced, are
within the error uncertainties of the respective parameters. In case of Stokes
I, the introduced error is slightly larger, peaking around 2 $\sigma$. This
enhanced sensitivity of I is due to the change of continuum intensity
introduced by gas pressure modifications.

\section{Results and Discussion}
\begin{figure*}
   \centering
   \includegraphics[angle=0,width=21cm]{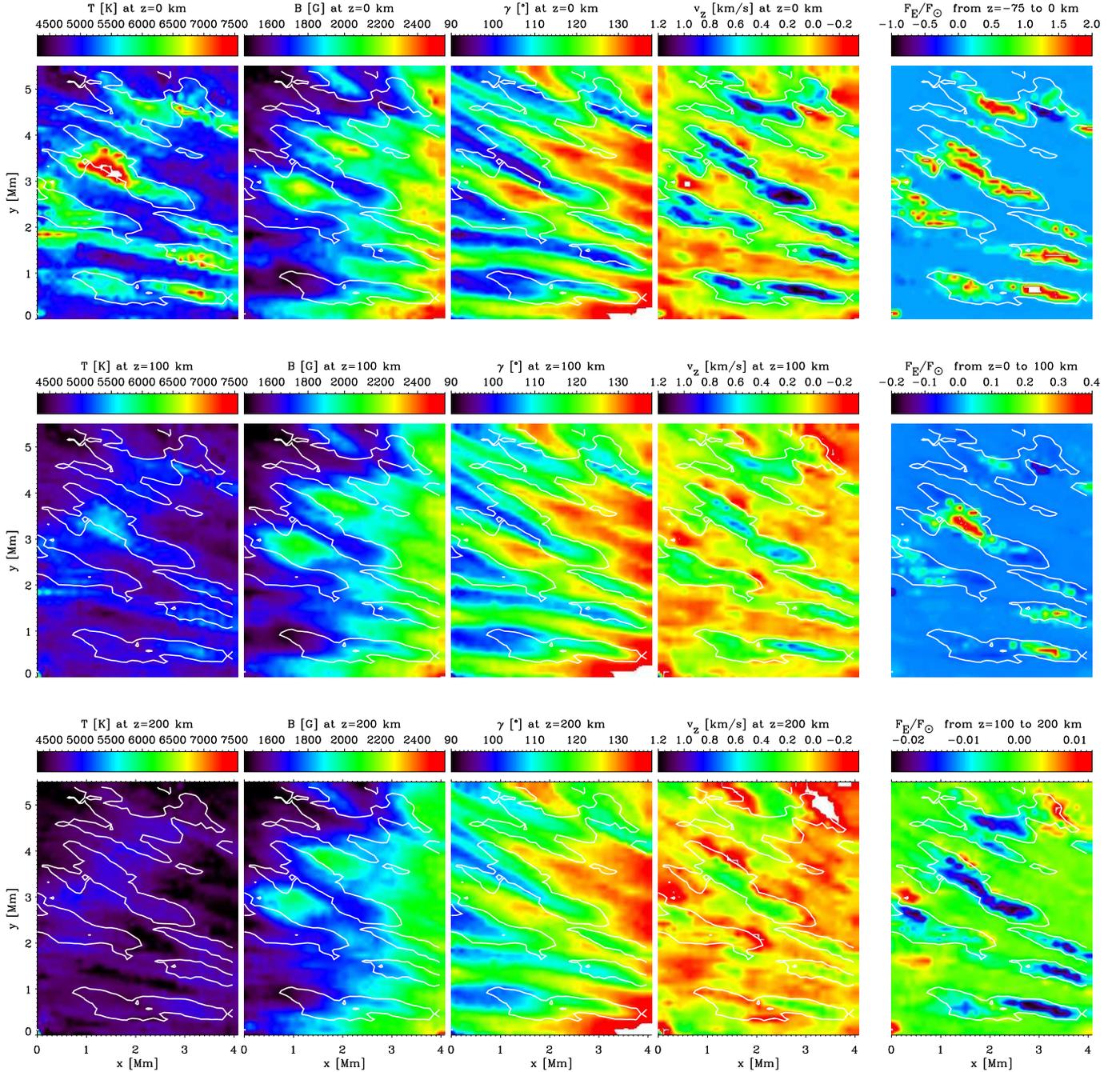}
   \caption{Maps of temperature $T$, magnetic field strength $B$, field
   inclination $\gamma$, and vertical component of velocity $v_{\rm z}$ at
   three different geometrical heights:$z$\,=\,0\,km (upper panels),
   $z$\,=\,100\,km (middle panels), $z$\,=\,200\,km (lower panels). The power
   per surface unity, $F_{\rm E}$, transferred by plasma motions in vertical
   direction between $z$\,=\,-75\,km and $z$\,=\,0\,km (upper right panel),
   between $z$\,=\,0\,km and $z$\,=\,100\,km (middle right panel), and between
   $z$\,=\,100\,km and $z$\,=\,200\,km (bottom right panel) is presented in
   units of the solar flux, respectively. White iso-contours enclose areas with
   velocities at the $z$\,=\,0\,km level larger than 0.3\,km\,s$^{-1}$ (strong upflows).}
   \label{paramphys}
\end{figure*}

In Figure~\ref{paramphys}, we present horizontal slices of different physical
quantities at three different height layers, i.e. $T$, $B$, $\gamma$, $v_{\rm
z}$, at 0\,km, 100\,km, and 200\,km, respectively (the most right panels of
each row will be described later). White contour lines correspond to $v_{\rm
z}$ larger than 0.3\,km\,s$^{-1}$ (i.e., relatively
strong upflows) as estimated at 0\,km. In all panels of this figure, the typical
filamentary structure of the penumbra is clearly observed, showing a
qualitatively good correlation between the different physical quantities. 
 The temperature contrast strongly diminishes with height, being nearly isothermal at $z$\,=\,200\,km; $\sigma
(T)$\,/\,$\overline{T}$ changes from 11\% at $z$\,=\,0\,km to 3.7\% at
$z$\,=\,200\,km. The mean magnetic field strength slightly diminishes from
1859\,G to 1794\,G in the same height range, while the mean field
inclination increases by 3$^{\circ}$. $\overline{v}_{\rm z}$ changes from
an upflow of 0.22\,km s$^{-1}$ to zero. Areas with ascending velocities larger 
than 0.3\,km\,s$^{-1}$ correspond to places with higher temperatures and weaker 
and more horizontal magnetic fields. 

\begin{table*}
\caption{Mean value and standard deviation of temperature $T$, magnetic field
strength $B$, field inclination $\gamma$, gas pressure $P_{\rm g}$, density
$\rho$, at three different height layers (0\,km, 100\,km, 200\,km) for points
harboring vertical velocities at $z$\,=\,0\,km ($v_{\rm 0}$) smaller or
larger than 0.3\,km\,s$^{-1}$, respectively.}
\begin{center}
\begin{tabular}{ccccccc}
\hline
 & Weak up/downflows&&&&& \\
$z$ [km] & Strong upflows & $T$ [K]      & $B$ [G]      & $\gamma$ [$^\circ$] & $P_{\rm g}$  [10$^{3}$ dyn\,cm$^{-2}$] &  $\rho$ [10$^{-7}$ g\,cm$^{-3}$] \\
\hline
         &$v_{\rm 0}$\,$<$\,0.3\,km\,s$^{-1}$ & 5175\,$\pm$\,471 & 1895\,$\pm$\,200 & 119\,$\pm$\,11     & 66\,$\pm$\,49     & 1.9\,$\pm$\,1.2  \\
    0    &$v_{\rm 0}$\,$>$\,0.3\,km\,s$^{-1}$ & 5769\,$\pm$\,614 & 1777\,$\pm$\,177 & 110\,$\pm$\,7      & 125\,$\pm$\,47    & 3.3\,$\pm$\,1.0  \\
\hline
         &$v_{\rm 0}$\,$<$\,0.3\,km\,s$^{-1}$ & 4802\,$\pm$\,203 & 1851\,$\pm$\,231 & 120\,$\pm$\,9      & 30\,$\pm$\,23     & 1.0\,$\pm$\,0.7  \\
  100    &$v_{\rm 0}$\,$>$\,0.3\,km\,s$^{-1}$ & 5001\,$\pm$\,204 & 1763\,$\pm$\,166 & 112\,$\pm$\,6      & 57\,$\pm$\,24     & 1.8\,$\pm$\,0.7  \\
\hline
         &$v_{\rm 0}$\,$<$\,0.3\,km\,s$^{-1}$ & 4557\,$\pm$\,172 & 1810\,$\pm$\,207 & 121\,$\pm$\,9      & 14\,$\pm$\,10     & 0.5\,$\pm$\,0.3  \\
  200    &$v_{\rm 0}$\,$>$\,0.3\,km\,s$^{-1}$ & 4695\,$\pm$\,123 & 1755\,$\pm$\,163 & 115\,$\pm$\,6      & 25\,$\pm$\,11     & 0.8\,$\pm$\,0.3  \\
\hline
\end{tabular}
\end{center}
\label{table1}
\end{table*}

\begin{figure}
   \centering
   \includegraphics[angle=0,width=9.cm]{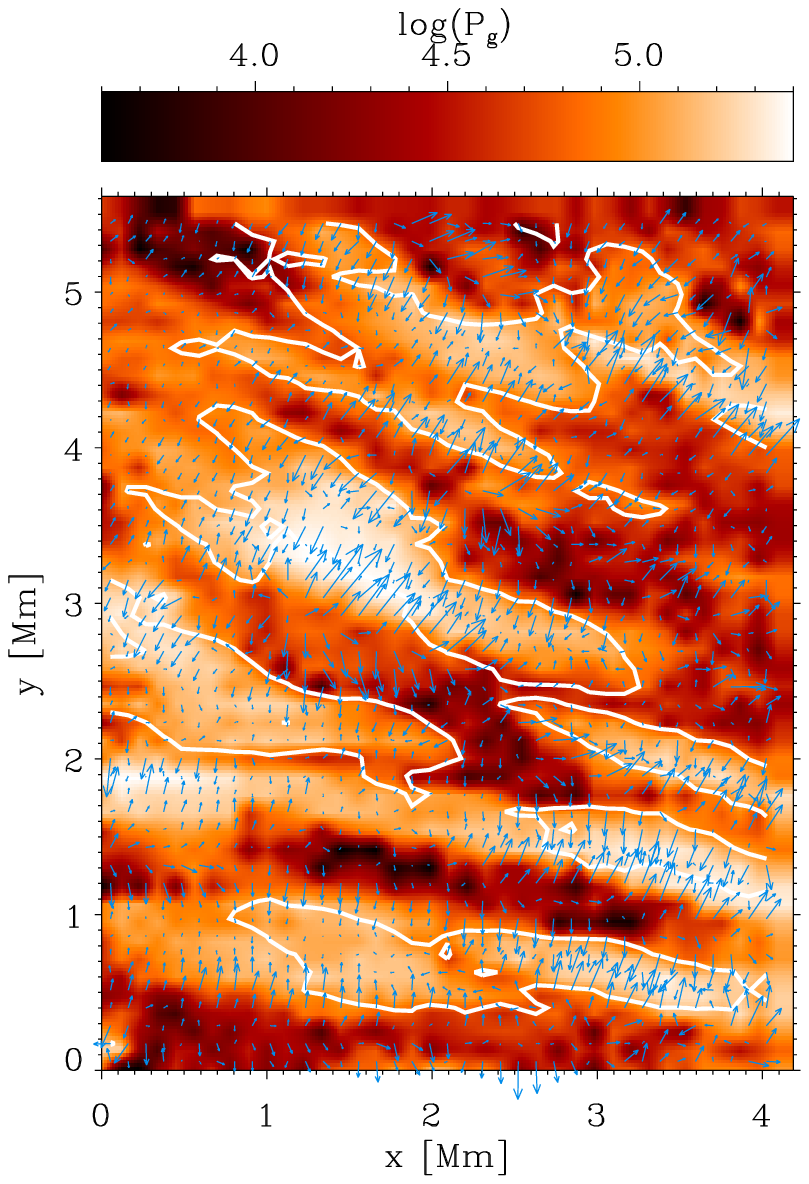}
   \caption{Map of gas pressure [dyn cm$^{-2}$] and horizontal component of the Lorentz force (blue arrows) at $z$\,=\,0\,km. White iso-contours are the same 
as in Figure~\ref{paramphys}.}
   \label{FigureNew}
\end{figure}

In Table~\ref{table1}, we summarize the mean value and standard deviation of some
quantities grouped in terms of vertical velocity at $z$\,=\,0. If we interpret the ascending 
velocities as the line-of-sight component of the Evershed flow, the behavior of $T$, ${V}_{\rm z}$, 
$B$ and $\gamma$ confirms the main properties of the uncombed scenario: ascending hotter material shows weaker and more horizontal magnetic fields.
 Interestingly, the upflowing material shows {\em enhanced gas pressure and densities} by around 86\% and 71\% respectively at all layers.
 Our results do not show any field-free region but rather indicate only a weak reduction of the magnetic field strength (from the nearly vertical background field of $\sim$\,1900\,G to the more horizontal field of bright penumbral filaments of $\sim$\,1750\,G) in the observed penumbral region.
The gradient of gas pressure, appearing at the borders of ascending areas, can not be compensated by the small variation of the magnetic pressure 
generated by the magnetic field weakening: it is balanced by the magnetic tension, which is produced by the strong variation of the magnetic field inclination. This is the reason why the method to establish an absolute geometrical height based on the equal total pressure between adjacent pixels \citep{san05} can produce erroneous solutions. In Figure~\ref{FigureNew} we show a map of the gas pressure at $z$\,=\,0; the contour lines, enclosing areas harboring strong upflows, are clearly related to areas with increased gas pressure. The horizontal component of the Lorentz force (blue arrows) nearly balances the gas pressure gradient.

The weak reduction of the magnetic field strength is in contradiction with the gappy model (Scharmer \& Spruit 2006 and Spruit \& Scharmer 2006) that predicts "nearly" field-free gaps harboring convective motions and reaching photospheric layers. A possible explanation could be a lack of spatial resolution: \cite{scharmeretal08} estimate that the minimum magnetic field strength in a dark cored penumbral filament is reduced by 25\% when the spatial resolution is increased from 0\farcs3 to 0\farcs15. Following this order of magnitude estimation, the strength of our more horizontal fields would reduce to $\sim$\,1300\,G, i.e. a figure still in the kG range like also found by \citet{borrerosolanki10}. The simulations of \cite{rempeletal09a, rempeletal09b} present reduced field strengths down to 700-1000\,G.

Besides, we do not find any signature of overturning convection in the penumbral area observed (at our resolution). 
 In Table~\ref{table2}, we present the net mass flow evaluated at three different geometrical heights:
0\,km, 100\,km, 200\,km. The uncertainties have been calculated after the introduction of an
artificial velocity offset of $\pm 100$\,m\,s$^{-1}$ considering possible uncertainties in the wavelength calibration. The average mass flow is strongly ascending and cannot be considered as the residual of the cancellation between ascending and descending parcels: At 0\,km and 100\,km, the upward directed mass flow is $\sim$\,5 times larger than the downward one and $\sim$\,2 times larger at 200\,km.
Our results appear to be compatible with mass motions dominated by the Evershed flow which are mainly upwards 
directed in the inner penumbrae \citep{rimmelemarino06}. In the panels of the 4$^{\rm th}$ column of Figure~\ref{paramphys}, 
the weak downflows almost exclusively appear in areas harboring stronger and more vertical (background) 
magnetic field. According to the gappy scenario, we should find downflows only at the borders of, but 
still inside, features with reduced and more horizontal magnetic fields. Perhaps, the lack of these downflows might be 
explained again by insufficient spatial resolution: \citet{zakharovetal08}, analyzing data obtained at the 
1-m Swedish Solar telescope, found weak downflows of around 90 m/s (certainly at the limit of the velocity accuracy) at the side of one filament. Although, it is questionable wether these weak downflows can be significant enough to counterbalance the factor 5 between  $<$\,$\rho v_{\rm z}$\,$>$\,$_{\rm{up}}$ and $<$\,$\rho v_{\rm z}$\,$>$\,$_{\rm{down}}$ found in the present work.

\begin{table}
\caption{Mean value of net mass flow over the full observed area ($<$\,$\rho v_{\rm z}$\,$>$),
mean value of mass flow in ascending  ($<$\,$\rho v_{\rm z}$\,$>$\,$_{\rm{up}}$) and descending ($<$\,$\rho v_{\rm z}$\,$>$\,$_{\rm{down}}$) areas.
 Mass flow given in units of [{\rm{g\,s}}$^{-1}$\,{\rm{m}}$^{-2}$]. Percentage of ascending areas (N[\%]).}
\begin{center}
\begin{tabular}{rcccc}
\hline
$z$ [km] & $<$\,$\rho v_{\rm z}$\,$>$ & $<$\,$\rho v_{\rm z}$\,$>$\,$_{\rm{up}}$ & $<$\,$\rho v_{\rm z}$\,$>$\,$_{\rm{down}}$ &  N\,[\%] \\
\hline
    0   & 72\,$\pm$\,23  &  96\,$\pm$\,8       & -17\,$\pm$\,4        & 78\,$\pm$\,22 \\
\hline 
   100  & 20\,$\pm$\,12  &  32\,$\pm$\,2       & \,\,\,-6\,$\pm$\,4   & 70\,$\pm$\,25 \\
\hline
   200  &  2\,$\pm$\,5   &  \,\,\,9\,$\pm$\,2  & \,\,\,-5\,$\pm$\,1   & 52\,$\pm$\,28 \\
\hline

\end{tabular}
\end{center}
\label{table2}
\end{table}

The uncombed penumbral model and the gappy model differ mainly in the way the energy is transferred. In the uncombed model the energy is carried by the Evershed flow \citep{Ruizcobobellotrubio08}, while the gappy model roots on convective energy transfer in field free regions \citep{scharmerspruit06}. For the first time we are able to estimate the energy flux stratification directly from observational data. The stratification of physical quantities in common geometrical height scale delivers all the ingredients needed for this calculation. In the right panels of Figure~\ref{paramphys} we show the energy flux (power per surface unity) $F_{\rm E}$ carried by the ascending material between two height layers in units of the solar flux $F_{\odot}$\,=\,$\sigma T_{eff}^{4}$. We have evaluated $F_{\rm E}$ by the integration of
\begin{equation}
\frac{\partial F_{\rm E}}{\partial z}=-\rho c_{\rm p} v_{\rm z} \left[\frac{dT}{dz}-\left(\frac{dT}{dz}\right)_{a}\right]\,\, ,
\label{FE}
\end{equation}
where $c_{\rm p}$ is the specific heat at constant pressure and the adiabatic temperature gradient is evaluated as
\begin{equation}
\left(\frac{dT}{dz}\right)_{a}=T\frac{d \log P_{\rm g}}{dz} \nabla_{a}\,\, ,
\end{equation}
with $\nabla_{a}$ is the double logarithmic isentropic temperature gradient. Both $c_{\rm p}$ and $\nabla_{a}$
have been evaluated taking into account partial ionization of hydrogen.


In the upper right panel of Figure~\ref{paramphys}, $F_{\rm E}$ integrated from
$z$\,=\,-75\,km to $z$\,=\,0\,km is depicted in units of solar flux. Only in areas with strong upflows 
(white contours), does the energy flux reach significant values and even exceed (in some points) the
solar flux up to 5 times (the right hand panels of Figure 10 are not 
scaled to the respective min/max value for contrast enhancement). A strong 
decrease of the energy flux with height is apparent. The contribution of the 0\,km to 100\,km layer to the energy flux
reaches maximum values equal to the solar flux (medium right panel), while $F_{\rm
E}$ integrated from $z$\,=\,100\,km and $z$\,=\,200\,km, (bottom right panel)
changes its sign, indicating that the ascending plasma is increasing its energy from the environment.

\begin{figure}
   \centering
   \includegraphics[angle=0,width=9cm]{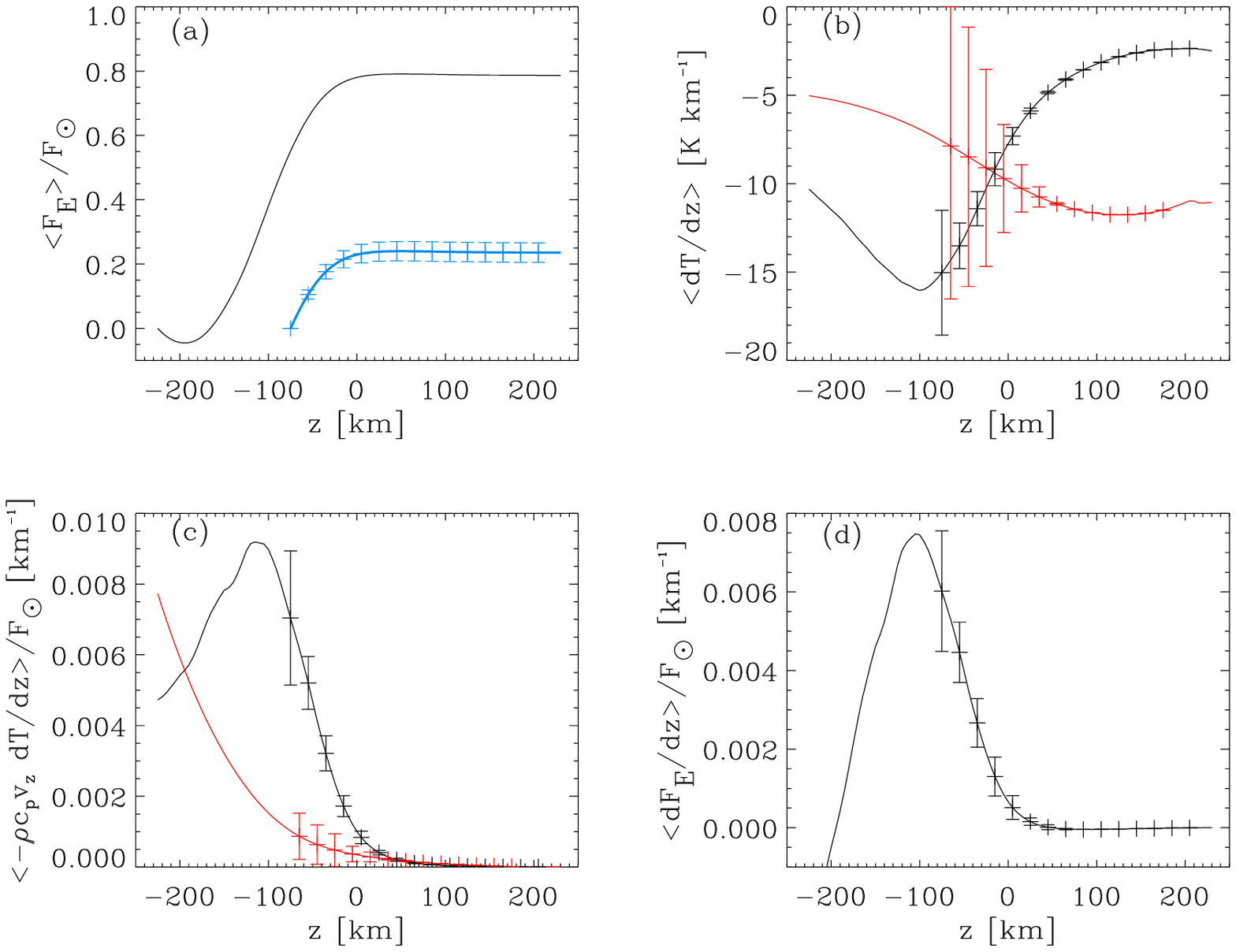}
   \caption{Panel (a): Energy flux $<$\,$F_{\rm E}$\,$>$ in solar flux units
   evaluated from the integration of the function plotted in panel (d) from 
   -225\,km until $z$. Contribution to the energy flux of layers higher than -75\,km
   together with their uncertainties (blue). Panel (b): mean temperature gradient (black) and mean
   adiabatic temperature gradient (red) averaged over the region under
   study. The error bars represent the errors of the mean temperature gradients
   above -75\,km. Panel (c): Height derivative of the energy flux carried by the
   vertical component of the plasma motion under constant pressure conditions
   $<$\,$-\rho c_{\rm p} v_{\rm z} \frac{dT}{dz}$\,$>$ in units of the solar
   flux (black), and the correction (red) in the height gradient of
   the energy flux which must be subtracted to account for pressure variations
   $<$\,$-\rho c_{\rm p} v_{\rm z} \left(\frac{dT}{dz}\right)_{a}$\,$>$. Panel
   (d): height derivative of the energy flux carried by the vertical component
   of the plasma motion in units of the solar flux.}
   \label{flujomedio}
\end{figure}

\begin{figure}
   \centering
   \includegraphics[angle=0,width=8cm]{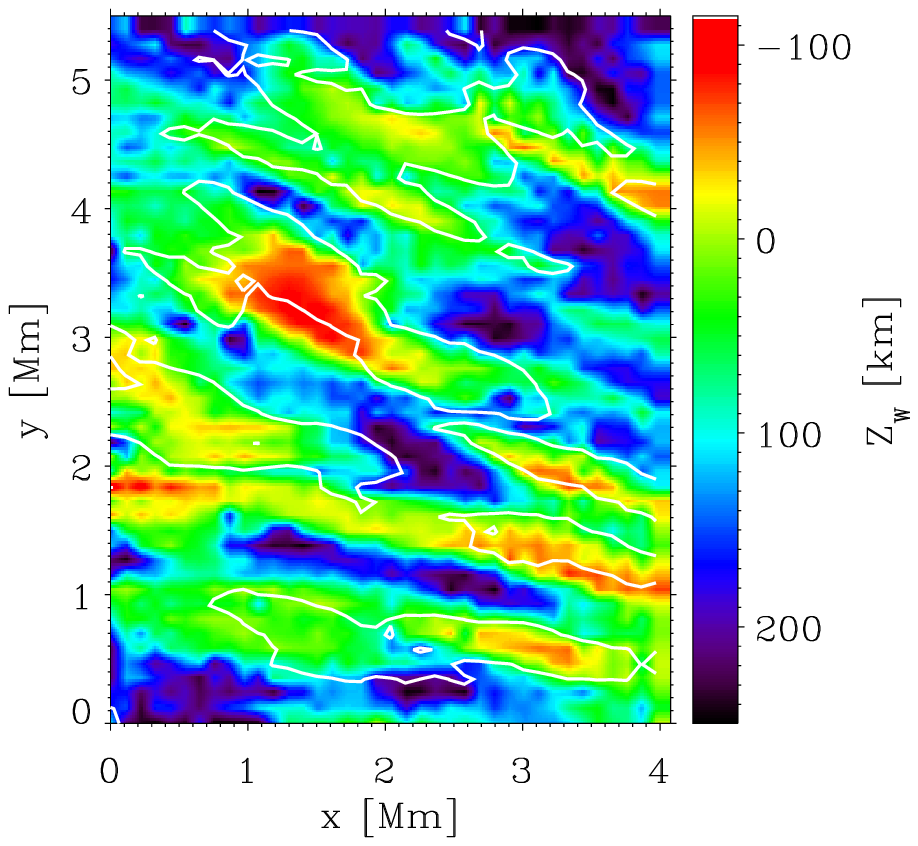}
   \caption{Map of Wilson depression $Z_{\rm W}$. White iso-contours enclose areas with
   velocities at $z$\,=\,0\,km larger than 0.3\,km\,s$^{-1}$ (strong upflows). The origin of the $Z_{\rm W}$-scale is arbitrary and $\sim$\,100\,km should be added.}
   \label{zw}
\end{figure}

Between $z$\,=\,-75\,km and $z$\,=\,0\,km the flux averaged over the area under
study amounts to 23\,\% of the solar flux, while the medium and upper layers
contribute with 0.9\,\% and -0.2\,\%, respectively. We determine the total amount of energy
flux carried by the ascending material in our full height range to be 24\,\%
which is well below the expected 75\,\% in sunspot penumbrae. Owing to the strong increase
of the energy flux with depth one can expect that the contribution of layers
below $z$\,=\,-75\,km might be sufficient to explain the observed brightness of
penumbra as channeled by the upflows.

As an order of magnitude estimate we have evaluated the contribution to the
energy flux $F_{\rm E}$ of the deep layers located between $z$\,=\,-225\,km and
$z$\,=\,-75\,km (for reasons that will become evident later). Clearly, the visible lines used in this work are nearly
insensible to physical parameters at layers deeper than -75\,km, i.e., the
stratification of temperature, density, pressure and velocity at such layers
result from linear extrapolations (in the original log($\tau$) scale); and so the
results at these layers can be considered only qualitatively. In panel (a) of
Figure~\ref{flujomedio} we plot the energy flux carried by the vertical motion
of the solar plasma between -225\,km and 200\,km, averaged over the penumbral
region studied here. The total contribution of the energy flux in the full
height range amounts to 78.7\% of the solar flux, a value near the
average brightness of a sunspot penumbra. The contribution to the energy flux of layers probed by the
visible lines is also plotted (in blue). At these layers, the error bars have been calculated by a Monte
Carlo simulation from the uncertainties of the physical quantities produced by
the inversion method and the genetic algorithm. In order to understand why
only layers between -180\,km and 20\,km appreciatively contribute to the
energy flux, we plot in panel (b) of Figure~\ref{flujomedio} the mean
temperature gradient (black) and the mean adiabatic temperature gradient (red) averaged over
the region under study. The strong increase of the uncertainties of these
gradients with depth reflects the above mentioned insensitivity of the lines
under study at deeper layers. Nevertheless, below 0\,km the temperature
gradient is seen to be steeper than the adiabatic one, thus at these
layers the ascending matter transfers energy to the surroundings. Apart from
temperature gradients, it is necessary to take into account density, vertical
velocity and specific heat stratification: in panel (c) of Figure~\ref{flujomedio},
we plot $<$\,$-\rho c_{\rm p} v_{\rm z}\frac{dT}{dz}$\,$>$ (black), i.e. the height derivative of the energy flux carried by
the vertical component of the plasma motion under constant pressure
conditions. The red line represents the correction in the gradient of energy
flux which must be subtracted to account for pressure variations $<$\,$-\rho
c_{\rm p} v_{\rm z} \left(\frac{dT}{dz}\right)_{a}$\,$>$. Under the caveat
mentioned above we can understand in the light of this figure why layers above
-200\,km are heated by the ascending material.
\begin{figure*}
   \centering
   \includegraphics[angle=0,width=13.0cm]{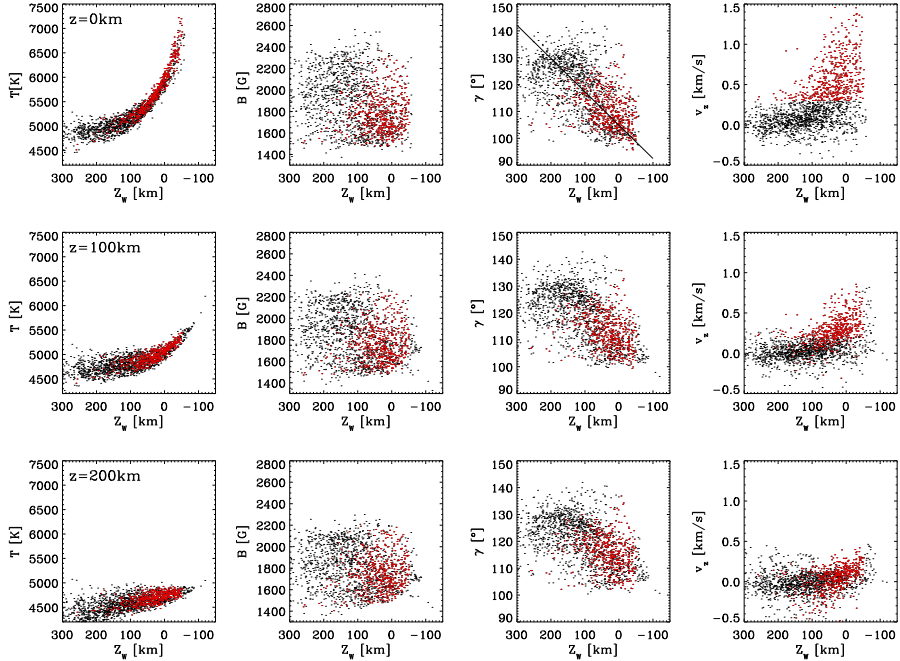}
   \caption{Scatter plots of temperature $T$, magnetic field strength $B$,
   field inclination $\gamma$, and vertical component of velocity $v_{\rm z}$
   {\it vs.} Wilson depression; at $z$\,=\,0\,km (upper panels),
   $z$\,=\,100\,km (middle panels), and $z$\,=\,200\,km (bottom panels). Points
   with vertical velocities at $z$\,=\,0\,km larger than 0.3\,km\,s$^{-1}$ (strong upflows) are
   presented in all panels by red colour. The origin of the $Z_{\rm W}$-scale is arbitrary and $\sim$\,100\,km should be added.}
   \label{scatter_zw}
\vspace{0.8truecm}
\end{figure*}

In panel (d) of Figure~\ref{flujomedio},
we present the height derivative of the energy flux carried
by the vertical component of the plasma motion in units of the solar flux
evaluated as the difference between the black and red line of panel (c), which is
essentially the average of equation~(\ref{FE}).


Another important parameter, which can be evaluated once the absolute
geometrical height scale is known, is the Wilson depression, i.e. the depth where the continuum optical depth equals 1. The knowledge of $Z_{\rm W}$ allows us to study which penumbral features are elevated or depressed. In Figure~\ref{zw}
we present the map of $Z_{\rm W}$.
Areas harboring significant up-flows (enclosed by the iso-velocity lines)
correspond to elevated structures ($Z_{\rm W}$\,$<$\,100\,km). When comparing
with Figure~\ref{paramphys}, a good correlation with the different physical
quantities is apparent. For a quantitative analysis of this correlation,
scatter plots of $T$, $B$, $\gamma$, $v_{\rm los}$ {\it vs.} the Wilson
depression are depicted in Figure~\ref{scatter_zw} for several geometrical heights.
 We observe a strong correlation between temperature $T$ and $Z_{\rm W}$: in
places with higher $T$, $\log(\tau)$\,=\,0 is displaced to upper layers due to
the strong dependence of opacity with temperature \citep[in agreement with the results found by][]{schmidtfritz04}.
 Note that, as stated above, the $z$\,=\,200\,km layer is nearly isotherm. Furthermore, for depressed
areas (e.g. $Z_{\rm W}$\,$>$\,150\,km) the difference in temperature between
the heights of 0\,km and 200\,km is only $\sim$\,500\,K, while for
elevated areas (e.g. $Z_{\rm W}$\,$<$\,0\,km), temperature differences of the
order of 2000\,K between both layers are reached. The magnetic field strength
$B$ shows a very weak correlation with $Z_{\rm W}$: lower $Z_{\rm W}$ tends to
be related to weaker magnetic fields. Nevertheless, we find clear trends
between field inclination $\gamma$ and the Wilson depression, at all layers.
 The same behavior is observed in case of the vertical component of velocity
$v_{\rm z}$. In Figure~\ref{paramphys} we have plotted the points corresponding
to pixels harboring significant upflows (in red). Focusing on the distribution of red
points throughout the panels, we can conclude that the Evershed flow corresponds
to areas with decreased Wilson depression (elevated structures), hotter temperatures, and weaker and
more horizontal magnetic fields. The mean value and the standard deviation of
the Wilson depression in areas with stronger ascending velocities ($v_{\rm
z}$\,$>$\,0.3\,km\,s$^{-1}$), amount to 34\,km\,$\pm$\,57\,km, while in the remaining areas values of
122\,km\,$\pm$\,82\,km are retrieved: i.e. features harboring the Evershed
flow are in average clearly elevated.

The zero in the $z$-scale (and consequently the zero of the Wilson depression) is arbitrary. Since the area where the genetic algorithm has been applied did not include the surroundings of the spot, an absolute scale with respect to the quiet
photosphere was not directly inferred from our analysis. However, such an absolute scale can be established in the following way. If we assume that the central part of the umbra has a Wilson depression of 600 km \citep{solanki03} with an inclination of 180 degrees, a linear continuation of the data points in the third panel of the upper row in Figure~\ref{scatter_zw} (see linear fit) establishes our mean umbral $\tau$\,=\,1 level at 500\,km, suggesting that our 0 km reference level is $\sim$\,100\,km depressed with respect to the normal quiet Sun. This indicates that our mean Wilson depression over the FOV (94\,$\pm$\,86\,km in our scale) corresponds to a depression of $\sim$\,194 km with respect to the surrounding quiet Sun atmosphere. This value is in agreement with previous estimates of the Wilson depression at the inner penumbra \citep{ruedi95}.

\section{Conclusions}

Using a genetic algorithm we have established a common geometrical height scale for photospheric penumbral models resulting from SIR inversions of Hinode data.
 Although the genetic algorithm only minimizes $\nabla\cdot\vec{B}$ and the deviations from static equilibrium
in the equation of motion at a height of 200\,km, the
resulting atmospheric model, interpolated to a common $z$\,-\,scale, shows very
small $\nabla\cdot\vec{B}$ values (and in any case, always inside the calculated
error bars) in a height range from 50\,km to 200\,km. The horizontal
derivatives of the divergence are of similar magnitude to that
of the vertical one. However, the latter has a mean value (0.2\,G/km)
that cannot be compensated by the mean values of the horizontal derivatives.
 This shows that, even at Hinode resolutions of 0.3\arcsec\, (200 km), the penumbra
displays unresolved structure. This unresolved structure can be present in both
horizontal and vertical directions. In the first case, improved spatial resolution
should reduce the mean value of the vertical derivative. In the second case, the
only way to achieve better resolution in the $z$-direction, would be by including more
spectral lines in the inversion. 

Having at our disposal a common geometrical height scale, we can evaluate e.g. electric currents (which
will be analyzed in a forthcoming paper), the Wilson depression, and in general
the three dimensional structure of penumbral features. The results show a
significant correlation between the different physical quantities. Owing to the visible photospheric lines used, we are only able to reliably 
probe the penumbra in the range from $\sim$\,0\,km to $\sim$\,200\,km. At deep
layers, strong spatial contrast in all parameters is found, whereas at the highest
layers both velocities and temperatures are nearly spatially homogeneous.
 Penumbral filaments with strong upflows have a weaker and more horizontal
magnetic field, are hotter and more dense and exhibit a $\tau$\,=\,1 surface displaced to higher layers.
 All these properties fit into the picture proposed by the uncombed scenario.

We point out that at the resolution of the Hinode/SP there is no trace of field-free regions in the penumbral volume retrieved by observations (pixel volume of $\sim$\,200$^3$\,km$^3$). Our inversions show kG field strength at all points inside this penumbral volume. We do not find traces of overturning convection in the inner penumbral area analyzed: the mass flow evaluated at three different height layers is clearly dominated by upflows,  the amount of ascending material being five times larger than the descending one. This results might certainly change when analyzing the mid and outer penumbra due to the presence of strong downflow patches as found by \citet{ichimotoetal07a}.

Besides, the observed upward motion, here identified as the vertical component of the Evershed flow, seems to carry enough energy to explain the brightness of the penumbra if the physical quantities below $z$\,=\,-75\,km are extrapolated from the results of the inversion.

As the method presented here has been applied so far only on a small penumbral region its extension on the entire sunspot penumbra, including the umbra and moat regions, will be addressed in a future work.\\

\acknowledgments
 This work has been supported by the Spanish Ministerio de Ciencia e Innovaci\'on through projects ESP 2006-13030-C06-01, AYA2007-65602, AYA2007-63881, and the European Commission through the SOLAIRE Network (MTRN-CT-2006-035484). The genetic algorithm has been kindly 
provided by E. P\'aez Ma\~n\'a. We thank C. Beck, R. Schlichenmaier, and S.~K. Solanki for fruitfull discussion. 



\clearpage


\begin{thebibliography}{}
  \bibitem[Balthasar et al.(2009)]{balthasaretal09} Balthasar, H., Bello Gonz\'alez, N., Collados, M., et al. 2009, in: Strassmeier, K.G., Kosovichev, A.G., Beckmann, J.E. (eds.), Cosmic Magnetic Fields: From Planets, to Stars and Galaxies, IAU Symp. 259, p. 665
  \bibitem[Beck(2006)]{beck06} Beck, C. 2006, Ph.D.Thesis, Albert-Ludwigs-University, Freiburg
  \bibitem[Beck(2008)]{beck08} Beck, C. 2008, \aap, 480, 825
  \bibitem[Bello Gonz\'alez \& Kneer(2008)]{bellokneer08} Bello Gonz\'alez, N. \& Kneer, F. 2008, \aap, 480, 265
  \bibitem[Bello Gonz\'alez et al.(2005)]{bellogonzalezetal05} Bello Gonz\'alez, N., Okunev, O. V., Dom\'inguez Cerde\~na, I., Kneer, F., \& Puschmann, K. G. 2005, \aap, 434, 317
  \bibitem[Bellot Rubio(2009)]{bellot09} Bellot Rubio, L., in Magnetic coupling between the Interior and the Atmosphere of the Sun, S.S. Hassan and  Rutten (eds.), ASP Ser., 2009, in press, arXiv:0903.3619
  \bibitem[Bellot Rubio et al. (2004)]{bellotrubioetal04} Bellot Rubio, L. R., Balthasar, H., \& Collados, M. 2004, \aap, 427, 319 
  \bibitem[Borrero(2007)]{borrero07} Borrero, J. M. 2007, \aap, 471, 967
  \bibitem[Borrero (2009)]{borrero09} Borrero, J. M. 2009, Sci. China Ser. G, 52, 1670
  \bibitem[Borrero et al.(2005)]{borreroetal05} Borrero, J. M., Lagg, A., Solanki, S. K., \& Collados, M. 2005, \aap, 436, 333
  \bibitem[Borrero et al.(2008)]{borrero08} Borrero, J. M., Lites, B. W., \& Solanki, S. 2008, \aap, 481, L13
  \bibitem[Borrero \& Solanki(2010)]{borrerosolanki10} Borrero, J. M. \& Solanki, S. K. 2010, \apj, 709, 349
  \bibitem[Borrero et al.(2006)]{borreroetal06} Borrero, J. M., Solanki, S. K., Lagg, A., Socas-Navarro, H., \& Lites, B. 2006, \aap, 436, 333
  \bibitem[Brummell et al.(2008)]{brummelletal08} Brummell, N. H., Tobias, S. M., Thomas, J. H., \& Weiss, N. O. 2008, \apj, 686, 1454
  \bibitem[Carroll \& Kopf(2008)]{carrolkopf08} Carroll, T. A. \& Kopf, M. 2008, \aap, 481, L37
  \bibitem[Denker et al.(2007)]{denkeretal07} Denker, C. Deng, N. Rimmele, T. R. Tritschler, A. Verdoni, A. 2007, Sol. Phys., 241, 411D
  \bibitem[Frutiger et al.(2000)]{frutigeretal00} Frutiger, C., Solanki, S. K., Fligge, M., \& Bruls, J. H. M. J. 2000, \aap, 358, 1109
  \bibitem[Gizon et al.(2009)]{gizon09} Gizon, L., Schunker, H., Baldner, C. S., et al. 2009, Space Sci. Reviews, 144, 249
  \bibitem[Heinemann et al.(2007)]{heinemann07} Heinemann, T., Nordlund, \AA., Scharmer, G. B., \& Spruit, H. C. 2007, \apj, 669, 1390
  \bibitem[Ichimoto et al.(2007a)]{ichimotoetal07a} Ichimoto, K., Shine, R. A., Lites, B., et al. 2007a, PASJ, 59, 593 
  \bibitem[Ichimoto et al.(2007b)]{ichimotoetal07b} Ichimoto, K., Suematsu, Y., Tsuneta, S., et al. 2007b, Science, 318,1597 
  \bibitem[Jurc\'ak \& Bellot Rubio(2008)]{jrcakbellotrubio08} Jurc\'ak, J. \& Bellot Rubio, L. 2008, \aap, 481, L17
  \bibitem[Jurc\'ak et al.(2007)]{jurcak07} Jurc\'ak, J., Bellot Rubio, L., Ichimoto, K., et al. 2007, PASJ, 59, 601
  \bibitem[Langhans et al.(2007)]{langhansetal07} Langhans, S., Scharmer, G. B., Kiselman, D., \& L\"ofdahl, M. G. 2007, \aap, 464, 763
  \bibitem[Lites et al.(1993)]{litesetal93} Lites, B. W., Elmore, D. F., Seagraves, P., \& Skumanich, A. 1993, \apj, 418, 928
  \bibitem[Lites et al.(2001)]{litesetal01} Lites, B. W., Elmore, D. F., \& Streander, K. V. 2001, ASPC, 236, 33L
  \bibitem[L\"ofdahl(2002)]{loefdahl02} L\"ofdahl, M. G. 2002, Proc. SPIE, 4792, 146
  \bibitem[Mathew et al.(2003)]{mathewetal03} Mathew, S. K., Lagg, A., Solanki, S. K., et al. 2003, \aap, 410, 695 
  \bibitem[Mart\'\i nez Pillet(2000)]{marpill00} Mart\'\i nez Pillet, V. 2000, \aap, 361, 734
  \bibitem[Mart\'\i nez Pillet et al.(2009)]{marpill09} Mart\'\i nez Pillet, V., Katsukawa, Y., Puschmann, K. G., \& Ruiz Cobo, B. 2009, \apj, 701, L79
  \bibitem[M\"uller et al.(2006)]{muelleretal06} M\"uller, D. A. N., Schlichenmaier, R., Fritz, G., \& Beck, C. 2006, \aap, 460, 925
  \bibitem[M\"uller et al.(2002)]{muelleretal02} M\"uller, D. A. N., Schlichenmaier, R., Steiner, O., \& Stix, M. 2002, \aap, 393, 305
  \bibitem[Puschmann et al.(2007)]{pus07} Puschmann, K. G., Kneer, F., Nicklas, H., Wittmann, A. D. 2007, in: Kneer, F., Puschmann, K. G., Wittmann, A.D. (eds.) Modern Solar Facilities - Advanced Solar Science, p. 45
  \bibitem[Puschmann et al.(2006)]{pus06} Puschmann, K. G., Kneer, F., Seelemann, T., Wittmann, A. D. 2006, \aap, 451, 1151
  \bibitem[Puschmann et al.(2005)]{pus05} Puschmann, K. G., Ruiz Cobo, B., V\'azquez, M., Bonet, J. A., \& Hanslmaier, A. 2005, \aap, 441, 1157
  \bibitem[Puschmann \& Sailer(2006)]{puschmannsailer06} Puschmann, K. G. \& Sailer, M. 2006, A\&A, 454, 1011
  \bibitem[Rempel et al.(2009a)]{rempeletal09a} Rempel, M., Sch\"ussler, M., \& Kn\"olker, M. 2009a, \apj, 691, 640
  \bibitem[Rempel et al.(2009b)]{rempeletal09b} Rempel, M., Sch\"ussler, M., Cameron, R.H., \& Kn\"olker, M. 2009b, Science, 325, 171
  \bibitem[Rimmele \& Marino(2006)]{rimmelemarino06} Rimmele, T. \& Marino, J. 2006, \apj, 646, 593
  \bibitem[Ruedi et al.(1995)]{ruedi95} Ruedi, I., Solanki, S. K., \& Livingston, W. 1995, \aap, 302, 543
  \bibitem[Ruiz Cobo \& Bellot Rubio(2008)]{Ruizcobobellotrubio08} Ruiz Cobo, B. \& Bellot Rubio, L. R. 2008, \aap, 488, 749
  \bibitem[Ruiz Cobo \& del Toro Iniesta(1992)]{ruizcobodeltoro92} Ruiz Cobo, B. \& del Toro Iniesta, J. C. 1992, \apj, 398, 375
  \bibitem[S\'anchez Almeida(1998)]{san98} S\'anchez Almeida, J. 1998, \apj, 497, 967
  \bibitem[S\'anchez Almeida(2005)]{san05} S\'anchez Almeida, J. 2005, \apj, 622, 1292
  \bibitem[S\'anchez Almeida et al.(1996)]{sanetal96} S\'anchez Almeida, J., Ruiz Cobo, B., \& del Toro Iniesta, J. C. 1996, \aap, 314, 295
  \bibitem[S\'anchez Cuberes et al.(2005)]{monica05} S\'anchez Cuberes, M., Puschmann, K. G., \& Wiehr, E. 2005, \aap, 440, 345
  \bibitem[Scharmer(2006)]{scharmer06} Scharmer, G. B. 2006, \aap, 447, 1111 
  \bibitem[Scharmer(2008)]{scharmer08} Scharmer, G. B. 2008, Phys. Scr., 133, 014015
  \bibitem[Scharmer(2009)]{scharmer09} Scharmer, G. B. 2009, Space Science Reviews, 144, 229
  \bibitem[Scharmer et al.(2002)]{scharmeretal02} Scharmer, G. B., Gudiksen, B. V., Kiselman, D., L\"ofdahl, M. G., \& Rouppe van der Voort, L. H. M. 2002, Nature, Volume 420, Issue 6912, pp. 151
  \bibitem[Scharmer et al.(2008)]{scharmeretal08} Scharmer, G. B., Narayan, G., Hillberg, T., et al. 2008, \apj, 689, L69
  \bibitem[Scharmer \& Spruit(2006)]{scharmerspruit06} Scharmer, G. B. \& Spruit, H. C. 2006, \aap, 460, 605
  \bibitem[Schlichenmaier(2009)]{sch09} Schlichenmaier, R. 2009, Space Science Reviews, 144, 213
  \bibitem[Schlichenmaier \& Collados(2002)]{sch02} Schlichenmaier, R. \& Collados, M. 2002, \aap, 381, 668
  \bibitem[Schlichenmaier et al.(1998a)]{sch98a} Schlichenmaier, R., Jahn, K., \& Schmidt, H. U. 1998a, \aap, 337, 897
  \bibitem[Schlichenmaier et al.(1998b)]{sch98b} Schlichenmaier, R., Jahn, K., \& Schmidt, H. U. 1998b, \apj, 493, 121
  \bibitem[Schlichenmaier et al.(2002)]{schetal02} Schlichenmaier, R., M\"uller, D. A. N., Steiner, O., \& Stix, M. 2002 \aap, 381, L77
  \bibitem[Schlichenmaier \& Solanki(2003)]{sch03} Schlichenmaier, R. \& Solanki, S. K. 2003, \aap, 411, 257
  \bibitem[Schmidt \& Fritz(2004)]{schmidtfritz04} Schmidt, W. \& Fritz, G. 2004, \aap, 421, 735
\bibitem[Solanki(2003)]{solanki03} Solanki, S. K. 2003, Astronomy and Astrophysics Review, 11, 153
\bibitem[Solanki \& Montavon(1993)]{solankimontavon93} Solanki, S. K. \& Montavon, C. A. P. 1993, \aap, 275, 283
\bibitem[Solanki \& R\"uedi(2003)]{solankirueedi03} Solanki, S. K. \& R\"uedi, I. 2003, \aap, 411, 249
  \bibitem[Spruit \& Scharmer(2006)]{spruitscharmer06} Spruit, H. C. \& Scharmer, G. B.  2006, \aap, 447, 343
  \bibitem[S\"utterlin et al.(2004)]{suetterlinetal04} S\"utterlin, P., Bellot Rubio, L. R., \& Schlichenmaier, R. 2004, \aap, 424, 1049
\bibitem[Title et al.(1993)]{titleetal93} Title, A. M., Frank, Z., Shine, R. A., Tarbell, T. D., Topka, K. P., Scharmer, G. \& Schmidt, W. 1993, \apj, 403, 780
\bibitem[Tritschler(2009)]{tritschler09} Tritschler, A. 2009, In: Proceedings of the Second Hinode Science Meeting, M. Cheung et al. (eds.), ASP Conf Series, in press, arXiv:0903.1300
\bibitem[Tsuneta et al.(2008)]{tsuneta08} Tsuneta, S., Ichimoto, K., Katsukawa, Y., et al. 2008, Sol. Phys., 249, 167
\bibitem[Vernazza et al.(1981)]{vernazzaetal81} Vernazza, J. E., Avrett E. H., \& Loeser, R. 1981, ApJS, 45, 635 (VAL-C)
\bibitem[van Noort et al.(2005)]{vannoortetal05} van Noort, M., Rouppe van der Voort, L., \& L\"ofdahl, M. G. 2005, Sol. Phys., 228, 191
\bibitem[Volkmer et al.(2007)]{volkmeretal07} Volkmer, R., von der L\"uhe, O., Kneer, F., et al. 2007, in: Kneer, F., Puschmann, K.G., Wittmann, A.D. (eds.) Modern Solar Facilities - Advanced Solar Science, p. 39
\bibitem[Weiss et al.(2004)]{weissetal04} Weiss, N. O., Thomas, J. H., Brummell, N. H., \& Tobias, S. M. 2004, \apj, 600, 1073
\bibitem[Westendorp Plaza et al.(2001a)]{westend01a} Westendorp Plaza, C., del Toro Iniesta, J. C., Ruiz Cobo, B., Mart\'\i nez Pillet, V., Lites, B. W. \& Skumanich, A. 2001, \apj, 547, 1130
\bibitem[Westendorp Plaza et al.(2001b)]{westend01b} Westendorp Plaza, C., del Toro Iniesta, J. C., Ruiz Cobo, B., \& Mart\'\i nez Pillet, V. 2001, \apj, 547, 1148
\bibitem[W\"oger \&  von der L\"uhe(2008)]{woegervdl08} W\"oger, F. \&  von der L\"uhe, O. 2008, Proc. SPIE, 7019, 46
\bibitem[Zakharov et al.(2008)]{zakharovetal08} Zakharov, V., Hirzberger, J., Riethm\"uller, T. L., Solanki, S. K., \& Kobel, P. 2008, \aap, 488, L17

\end{thebibliography}
\end{document}